# Heisenberg Spin-1/2 Antiferromagnetic Molecular Chains

Kewei Sun[1,2], Nan Cao[3], Orlando J. Silveira[3], Adolfo O. Fumega[3], Fiona Hanindita[4], Shingo Ito[4]*, Jose L. Lado[3], Peter Liljeroth[3], Adam S. Foster[3,5]*, Shigeki Kawai[2,6]*

[1]*International Center for Young Scientists, National Institute for Materials Science, 1-2-1 Sengen, Tsukuba, Ibaraki 305-0047, Japan.*

[2]*Center for Basic Research on Materials, National Institute for Materials Sciences, 1-2-1 Sengen, Tsukuba, Ibaraki 305-0047, Japan.*

[3]*Department of Applied Physics, Aalto University, Espoo, Finland.*

[4]*Division of Chemistry and Biological Chemistry, School of Physical and Mathematical Sciences, Nanyang Technological University 21 Nanyang Link, 637371, Singapore.*

[5]*WPI Nano Life Science Institute (WPI-NanoLSI), Kanazawa University, Kakuma-machi, Japan.*

[6]*Graduate School of Pure and Applied Sciences, University of Tsukuba, Tsukuba 305-8571, Japan.*

**Abstract**

Carbon-based nanostructures possessing π-electron magnetism have attracted tremendous interest due to their great potential for nano spintronics. In particular, quantum chains with magnetic molecular units synthesized by on-surface reactions provide an ideal playground for investigating magnetic exchange interactions between localized spin components. Here, we present an extensive study of antiferromagnetic nanographene chains with the diazahexabenzocoronene molecule as the repeating unit. A combination of bond-resolved scanning tunneling microscopy, density functional theory and quantum spin models revealed their detailed structures and electronic and magnetic properties. We found that the antiferromagnetic chains host a collective state featuring gapped excitations for an even number of repeating units and one featuring a Kondo excitation for an odd number. Comparing with exact many-body quantum spin models, our molecular chains provide the realization of an entangled quantum Heisenberg model. Coupled with the tuneability of the molecular building blocks, these systems can act as an ideal platform for the experimental realization of topological spin lattices.

**Introduction**

Atomically defined carbon nanostructures hosting π-electron magnetism[1] have great potential as nanoscale magnetic components in spin-electronic devices[2,3,4]. Compared to *d/f*-electron magnetism in transition metal elements, π-electrons in the molecules are usually more delocalized and the weak spin-orbit interaction with long spin coherence time and length[5,6,7,8,9] is expected to exhibit tunable quantum properties. A key development target is the fabrication of various atomically defined carbon nanostructures with spin polarization, offering a platform for investigating low energy excitations. However, since molecules with unpaired electrons are usually highly reactive and short-lived, the synthesis is challenging by solution-based chemical reactions. On-surface synthesis has been developed to fabricate atomically accurate carbon-based nanostructures *via* activation and conjugation of precursor molecules on metal substrates[10,11], and has opened an exciting possibility of studying low energy spin excitations in carbon-based materials. Several nanographene (NG) structures and graphene nanoribbons (GNRs) with unpaired π-electrons or zigzag edges have been successfully synthesized and characterized on surfaces, and their magnetic properties give rise to a variety of correlated phases[12,13,14,15,16,17,18,19,20,21,22,23,24]. For example, magnetic chains have been fabricated by conjugating triangulenes and porphyrin oligomers, and the Coulomb repulsion between the π electrons in each unit that compose the chains gives rise to an antiferromagnetic ordering, bringing the whole chain to a topological Haldane spin chain phase[24,25,26,27]. Another strategy to achieve different correlated scenarios in a controlled manner is to consider nitrogen doping into the organic oligomers, creating a surplus of π electrons that populate previously unoccupied energy levels and bringing the whole system to completely different phases[28,29].

In this article, we demonstrate the realization of a text-book spin S = 1/2 Heisenberg chain (Scheme1)[30,31], within metal-free organic chains through a strategic on-surface synthesis of nitrogen doped chains derived from an aza-coronene building block. We synthesized a unit of the diazahexabenzocoronene ($N_2HBC$) molecule[32] and $N_2HBC$ chains that present a perfect alignment of spin S = 1/2 units represented by the simple Heisenberg Hamiltonian with exchange coupling *J* between neighboring units:

$$\hat{H} = \sum_i J \vec{S}_i \cdot \vec{S}_{i+1} \qquad (1)$$

where $\vec{S}_i$ denotes the spin half operator at site $i$. We were able to theoretically and experimentally establish the behavior of the Heisenberg chains and correlate the number of units that each chain is formed (even or odd) with their collective quantum spin excitations in low bias scanning tunneling spectroscopy (STS) experiments. These molecular chains offer a flexible platform for the design and experimental realization of spin lattices.

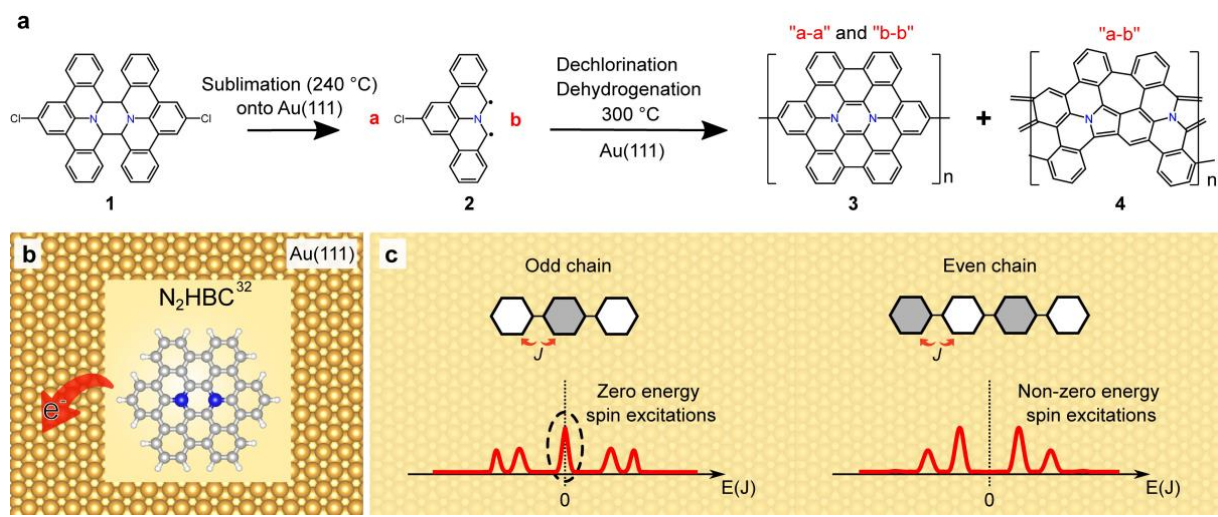

**Scheme 1. a**, Reaction processes to form an S = 1/2 unit and the oligomers (N$_2$HBC chains) as well as graphene nanoribbon. **b**, N$_2$HBC single molecule donates one electron to the surface, achieving an S = 1/2 spin state. **c**, Schematic drawing of Heisenberg S = 1/2 antiferromagnetic molecular chains composed of odd and even numbers of molecular units, and the corresponding magnetic couplings.

## Results and discussions
### On-surface synthesis of N$_2$HBC molecule and respective chains

The precursor molecule **1** was prepared according to the same procedure as in the literature (See Supplementary Information for the details)[33]. Annealing the precursor molecules on Au(111) at 300 °C led to the formation of the hexagonal monomer (dashed circles in Fig. 1a) and its oligomers with different lengths (yellow arrows in Fig. 1a), as well as nanoribbons (blue arrows in Fig. 1a). The synthesis of the nanoribbons indicates that the precursor molecule was initially split (molecule **2**) and subsequently connected to each other (see supplementary information, inset III of Fig. 1a, Supplementary Fig. S1-4). Since we are interested in the exploration of possible molecular spin structures, we focus on the hexagonal monomer and its oligomers, and do not discuss the nanoribbons further. To investigate the chemical structures of the products, the scanning tunneling microscopy (STM) tip apex was terminated with a carbon monoxide (CO) molecule[34,35]. Inset I of Fig. 1a shows the close-up view of the monomer, whose inner structure was clearly observed in the constant height d$I$/d$V$ map (Fig. 1b) and the corresponding Laplace filtered image (Fig. 1c). We found that the synthesis of the $N_2$HBC molecule (Fig. 1d), resulted from the C-C recoupling, dechlorination and dehydrogenation of two **2**. The constant height d$I$/d$V$ map taken near the Fermi level has a higher intensity at the center, suggesting that the net spin arises from a pair of N atoms in the monomer. Density functional theory (DFT) calculations indicate that the gas-phase $N_2$HBC molecule is in a closed-shell singlet state in its neutral form but assumes a cationic state when deposited on Au(111) due to misalignment of their Fermi levels, thus bringing the system to the open-shell $S = 1/2$ state. It was found that each $N_2$HBC unit approximately donates one electron until the Fermi level of the chains matches that of the underlying substrate (see the work function calculations in Supplementary Table S1). A simulated constant-height d$I$/d$V$ map of the positively charged $N_2$HBC$^+$ at 0 mV (Fig. 1e) captures all the main features of the experiment. Therefore, the central high-intensity signal is attributed to the Kondo resonance, which can be seen in the low bias d$I$/d$V$ spectrum (Fig. 1f). Inset II in Fig. 1a was a trimer-$N_2$HBC$^{3+}$, in which each unit is connected by a C-C bond (Fig. 1g,h). Such coordinated fusion leads to the structure seen in Fig. 1i possessing all nitrogen atoms along the longitudinal direction of the chain. Interestingly, the unit in the center of the trimer-$N_2$HBC$^{3+}$ chain shows a darker contrast than the other peripheral two units in the d$I$/d$V$ map (Fig. 1g).

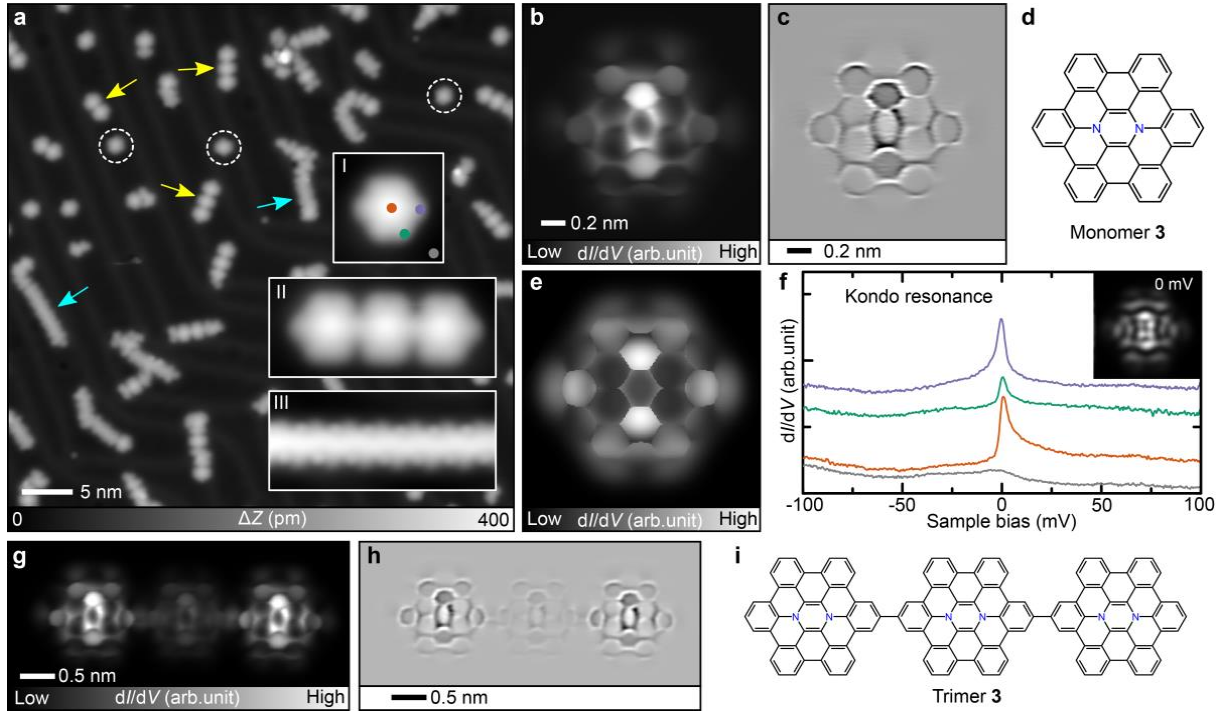

**Fig. 1. Synthesis of N$_2$HBC molecule and respective chains 3. a**, STM topography of a Au(111) surface after annealing at 300 °C for 5 min. Inset I, II, III show close-up views of three different products. **b**, Constant height d$I$/d$V$ map of individual **3** and **c**, the corresponding Laplace filtered image, and **d**, the chemical structure. **e**, The simulation of constant height d$I$/d$V$ map is performed using a relaxed CO tip. Constant height d$I$/d$V$ simulations were acquired with a mixture of $s$ and $p_{xy}$ wave tips considering different ratios. The comparison of simulated images with different tips is presented in Supplementary Note 1 and Supplementary Fig. S5,S6. **f**, d$I$/d$V$ curves recorded at three different sites on the molecule and one site on Au(111) as a reference, as indicated by colored dots in the left inset. The right inset shows the spatial distribution of the Kondo resonance state. **g,h,i**, Constant height d$I$/d$V$ map of individual trimer-N$_2$HBC$^{3+}$ (**g**) and the corresponding Laplace filtered image (**h**) and the chemical structure (**i**). Measurement parameters: Sample bias voltage $V$ = 200 mV and tunneling current $I$ = 5 pA in **a**. $V$ = 200 mV and $I$ = 5 pA in inset I of **a**, $V$ = 200 mV and $I$ = 3 pA in inset II of **a**, and $V$ = 200 mV and $I$ = 10 pA in inset III of **a**. $V$ = 1 mV and $V_{ac}$ = 10 mV in **b,g**.

**Electronic structure characterization of the N$_2$HBC chains**

We conducted STS measurements over a pentamer of N$_2$HBC (inset of Fig. 2a) and performed DFT and mean-field Hubbard calculations of a periodic chain made of $N_2HBC^+$ units. The d$I$/d$V$ spectra, along with that on a bare Au(111) substrate (grey curve) as a reference, exhibits several characteristic peaks (Fig. 2a). We identified –0.7 V as the singly occupied molecular orbital (SOMO in Fig. 2b) and 0.5 V as the singly unoccupied molecular orbital (SUMO in Fig. 2c), leading to an SOMO-SUMO gap of 1.2 eV (see Supplementary Fig. S7 for the electronic properties of the pentamer). The electronic properties of the $N_2HBC$ molecule and chains from dimer to tetramer were also measured by STS (Supplementary Fig. S8-S11), exhibiting similar SOMO-SUMO gaps and unoccupied states as compared to penta-$N_2HBC^{5+}$. Significant spectroscopic signatures observed around the Fermi level as indicated by the dashed rectangle in Fig. 2a suggest the presence of low energy spin excitations in the penta-$N_2HBC^{5+}$. Both DFT and mean-field Hubbard calculations of free-standing periodic chain consisting of dimer-$N_2HBC^{2+}$ in its unit-cell indicate that its ground state is antiferromagnetic. Mean-field Hubbard calculations with $U = 1.5t$ ($t$ is the nearest neighbors hopping parameter, we use $t$ = 2.7 eV) leads to the total energy difference between antiferromagnetic and ferromagnetic states of about 10 meV. This value depends on $U$, reaching up to 21 meV for $U = 2.0t$. Following the same trend, DFT-B3LYP total energies indicate that the antiferromagnetic periodic chain is 97 meV more stable than its ferromagnetic counterpart. The stability of the chains through their total energies can be used to estimate the exchange coupling $J$, since the spin excitations will bring the periodic chains to a ferromagnetic excited state from an antiferromagnetic ground state. The fact that the DFT-B3LYP and mean-field Hubbard modelling predict antiferromagnetic coupling between the units of the dimer-$N_2HBC^{2+}$ chains in the absence of a substrate suggests that the role of any RKKY interaction mediated by the metallic surface is negligible. The magnitude and nature of the coupling within the chains is more characteristic of the superexchange interaction seen in previous molecular systems[36,37,38]. In the next section we will discuss in more detail the comparison with the experimental data.

Figure 2h shows the DFT-B3LYP band structures and density of states (DOS) of the dimer-$N_2HBC^{2+}$ periodic chain, revealing a band gap of 1.1 eV, very close to the measured SOMO-SUMO gap of 1.2 eV (also seen Supplementary Note 2 and Supplementary Figure S12 and S13). The calculated LDOS maps obtained for the valence and conductance bands (see insets in Fig. 2h) highlights the antiferromagnetic behavior of the dimer-$N_2HBC^{2+}$ chain: both valence and conduction bands have alternating spin contributions in the central hexagon of the subsequent $N_2HBC^+$ units. Figure 2d-g show the constant current d$I$/d$V$ maps at corresponding sample bias demonstrating the evolution of spatial distribution for these unoccupied states, and the same trend of the inner $N_2HBC^+$ units is observed in the simulated d$I$/d$V$ maps in Fig. 2i-n. We note here that in the comparison between experimental (Fig. 2b-g) and simulated maps (Fig. 2i-n), the simulations are periodic and do not demonstrate the chain end effects seen in experiments.

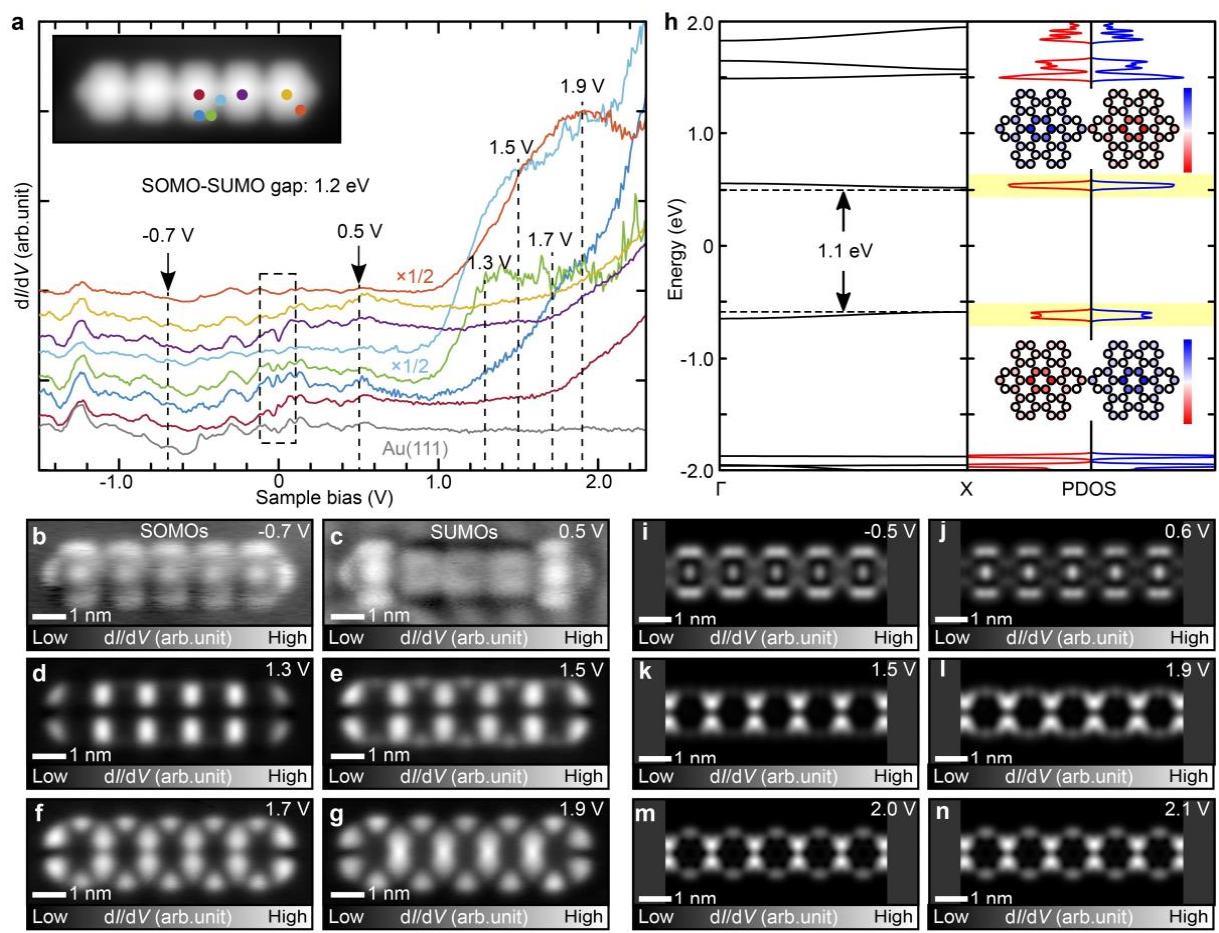

**Fig. 2. Electronic properties of pentamer-$N_2HBC^{5+}$ 3. a**, d$I$/d$V$ curves collected at different sites of the pentamer-$N_2HBC^{5+}$, as indicated by colored dots in its STM topography (inset). **b-g,** Constant current d$I$/d$V$ maps measured at sample bias voltages of –0.7, 0.5, 1.3, 1.5, 1.7, and 1.9 V, respectively. **h**, Band structure and spin polarized density of states (DOS) of a periodic antiferromagnetic chain. The inset shows the DOS map of the valence band (bottom) and conduction band (top) resolved by spin. **i-n,** Constant height d$I$/d$V$ maps were calculated at bias voltages corresponding to the simulated DOS. The simulations were carried out with a flexible CO tip (a combination of 13% of $s$ and 87% of $p_{xy}$ waves), which is comparable to the experimental ones. Measurement parameters: $V = -700$ mV, $I = 160$ pA and $V_{ac} = 10$ mV in (**b**). For STS measurement, the tip-sample distance was adjusted at the corresponding bias voltages and $I = 160$ pA, $V_{ac} = 10$ mV in (**c-g**).

**Low-bias measurements and an effective model for N₂HBC chains**

Next, we investigated the magnetic properties of the N₂HBC⁺ chains up to maximum observed experimentally, the pentamer, by low-bias constant height STS measurements around the Fermi level, in which we collected d$I$/d$V$ curves on the N₂HBC⁺ units of each chain and d$I$/d$V$ maps on selected bias values (Fig. 3). Local spin excitations that influence the inelastic tunneling spectroscopy of each chain were treated by the dynamical spin correlator (DSC) $\boldsymbol{A}(\omega, n)$:

$$\boldsymbol{A}(\omega, n) = \langle GS | S_n^- \delta(\omega + E_{GS} - \hat{H}) S_n^+ | GS \rangle, \qquad (2)$$

which takes the many-body ground state $|GS\rangle$ and energies $E_{GS}$ obtained by solving Equation 1[39,40,41]. The frequency of the excitation $\omega$ corresponds to the applied bias voltage in the experiment, $n$ indicates which unit is treated within the chain and $S_n^\pm = S_n^x \pm iS_n^y$. As the spin spectral weight of the magnetic units is related to the d$I$/d$V$ line shape[42], and consequently its spatial distribution we coupled the DSC to both DFT d$I$/d$V$ simulated maps and mean-field Hubbard spin densities by modulating their spatial distribution with $\boldsymbol{A}(\omega, n)$, allowing us to emulate the conduction maps and stablish their behavior with the size and parity of the chains and to their spin polarization.

We observed a systematic behavior of the d$I$/d$V$ curves and the DSC relative to the number of N₂HBC⁺ units, in which we assigned as even (Fig. 3a-l) or odd (Fig. 3m-x) Heisenberg chains according to their parity. For the dimer-N₂HBC²⁺ in Fig. 3a, there is no prominent Kondo peak at the Fermi level, and we find symmetric conduction peaks around the Fermi level in the d$I$/d$V$ curves acquired above each unit of the dimer. The constant height d$I$/d$V$ map of dimer-N₂HBC²⁺ collected at 1 mV reveals that the signal is indeed quite uniform over all the structure (Fig. 3c), while still resembling the signal from a single N₂HBC⁺ molecule at 0 V. The tunneling conduction peaks of the dimer-$N_2HBC^{2+}$ are consistent with the peaks of the DSC calculated with $J$ = 36 meV (Fig. 3b), in agreement with the experimental value (Supplementary Fig. S14) and within the range of our mean-field Hubbard estimated exchange coupling (between 20 and 40 mV) and DFT (see Supplementary Information for more details of the calculations, Supplementary Note 3 and Supplementary Fig. S12,S13,S15-17). The two spin 1/2 subunits in the dimer are antiferromagnetically coupled and the low-energy spectroscopic features correspond to an inelastic singlet-triplet excitation (S=0 to S=1). Both $N_2HBC^+$ units display equal intensity as simulated in the energy-normalized conductance maps, and both DFT and MF-Hubbard spin densities are concentrated over the central part of each molecule-unit where the N pair is located Fig. 3d. The same behavior as for the dimer is observed in the d$I$/d$V$ curves of the tetramer-$N_2HBC^{4+}$ (Fig. 3g), with no prominent Kondo peak and symmetric conduction peaks around the Fermi level, and the assignment of the phenomenology is similar. No strong feature is observed at 0 V in the d$I$/d$V$ curves, and the d$I$/d$V$ map taken at 1 mV has a slightly brighter signal at the peripheral $N_2HBC^+$ units (Fig. 3i). We then extended our dimer-$N_2HBC^{2+}$ model to a four spin 1/2 system in the tetramer-$N_2HBC^{4+}$, keeping the same exchange coupling $J$ = 36 mV. Indeed, the DSC gives several peaks around the Fermi level, consistent with the conduction peaks in the d$I$/d$V$ curves (Fig. 3h). Surprisingly, the DSC-modulated conductance maps simulations in Fig. 3j correctly display the slightly stronger intensity over both peripheral units in the tetramer-$N_2HBC^{4+}$ as observed in the experimental map. The MF-Hubbard spin density of the tetramer-$N_2HBC^{4+}$ shows that the spin distributes at center of each unit, similar to the one obtained for the dimer-$N_2HBC^{2+}$ (the right panel of Fig. 3j). Constant height conduction maps for both dimer and tetramer at higher energies display comparable spin intensity with the simulated ones (Fig. 3e,f and k,l), which also correlate to DSC-modulated simulations.

The spectroscopy features of the odd Heisenberg chains (Fig. 3m,s) seem to contrast to the features observed in the even cases, as several peaks with different intensities and broadenings are observed in the tunneling conductance curves at the Fermi level. However, we demonstrate that they are still characteristic of a chain with only exchange coupling between spin 1/2 $N_2HBC^+$ units. In Fig. 3m,s, the d$I$/d$V$ curves of both trimer-$N_2HBC^{3+}$ and pentamer-$N_2HBC^{5+}$ chains reveal zero-bias resonance peaks at the peripheral $N_2HBC^+$ units, with extra symmetric peaks in the d$I$/d$V$ curves at higher bias voltages. While the d$I$/d$V$ curves taken at the central $N_2HBC^+$ unit of the trimer-$N_2HBC^{3+}$ are apparently featureless at the Fermi level (Fig. 3m), the central $N_2HBC^+$ unit in the pentamer-$N_2HBC^{5+}$ has the same zero-bias features as the peripheral ones, with stronger intensity (Fig. 3s). In contrast, the $N_2HBC^+$ units next to the central one have no features at the Fermi level (see more details in the two-dimensional d$I$/d$V$ map composed of 121 d$I$/d$V$ curves taken along the pentamer-$N_2HBC^{5+}$ chain in Supplementary Fig. S18). The relative contrast of the molecular units inverts at higher bias - this is particularly visible in the d$I$/d$V$ maps: the constant height d$I$/d$V$ maps of the trimer-$N_2HBC^{3+}$ taken at 1 mV (Fig. 3o) corroborate to the strong intensity of the zero-bias peak at the peripheral $N_2HBC^+$ units, while at -30 mV (Fig. 3q) the signal from the central $N_2HBC^+$ unit becomes more intense relative to the peripheral units. The DSC calculated at the peripheral units of the trimer-$N_2HBC^{3+}$ (Fig. 3n) reveals a sharp peak at the Fermi level, consistent with the zero-bias resonances observed in the experiment, which is attributed to an emergent collective two-fold degenerate ground state that is only possible due to the odd number of spin 1/2 units, allowing the development of a Kondo resonance when in contact with the conduction electrons in the substrate. For higher energies, other steps consistent with higher order inelastic spin-excitations are also observed. On the other hand, the DSC calculated at the central unit of the trimer-$N_2HBC^{3+}$ reveals a quenched zero-bias peak, and a single sharp peak for higher and lower energies. Given the simplicity of the trimer-$N_2HBC^{3+}$ case, we were able to describe in more detail the zero-bias spin excitations by analytically solving the Hamiltonian in Equation 1. The two-fold degenerate many-body states that result in $S^z = \pm 1/2$ are: $|\alpha\rangle = \frac{1}{\sqrt{6}}(|\uparrow\uparrow\downarrow\rangle - 2|\uparrow\downarrow\uparrow\rangle + |\downarrow\uparrow\uparrow\rangle)$ and $|\beta\rangle = \frac{1}{\sqrt{6}}(|\downarrow\downarrow\uparrow\rangle - 2|\downarrow\uparrow\downarrow\rangle + |\uparrow\downarrow\downarrow\rangle)$ (See Supplementary Note 4 for the details). Note that the units at the edges of the chain are equivalent as given by mirror symmetry, whereas the central unit is different. This stems from the open boundary conditions of the chain. At zero frequency (or zero bias), the DSC is proportional to the $\langle\alpha|S_n^+|\beta\rangle^2$ matrix element of the ladder operator[42] that will flip a single spin in the unit $n$ of the trimer where the excitation occurs. The matrix element calculated flipping the spin in the peripheral units are four times larger than the one calculated at the central unit, giving rise to more intense zero-bias peak as seen in the DSC curve Fig 3n (see Supplementary Information for a short summary of the calculation). The DSC-modulated conductance map shown in Fig. 3p correctly displays the contrast obtained in the experiment, demonstrating that the zero-bias contrast is related to the development of a Kondo resonance when in contact with the conduction electrons in the substrate. The inverted behavior of the tunneling conductance at -30 mV is also captured in the simulated maps, which agrees well with the experimental one, shown in Fig. 3q,r.

Extending our analysis from the trimer-$N_2HBC^{3+}$ to the pentamer-$N_2HBC^{5+}$ becomes then straightforward with our model, reproducing especially all the zero-bias resonances and their respective intensities (see Fig. 3s-x), which are also related to elastic spin-excitations, and some inelastic excitations. Supplementary Fig. S19 shows a schematic visualization of the DSC calculated at several different energies from 1 to 45 meV, depicting the expected relative behavior of the contrasts in each unit for all chains. Together with that, it also shows that scaling from the numbers of molecules will allow us to explore the development of 1/2 Heisenberg spin structure according to the number of molecular components and the transition from expected Kondo resonances for odd chains to quenching in longer chains. The fine tuning of the spin excitation together with the full control of the number of molecules are key steps to engineering of spin coherence times and realizable spintronics. Additional supporting measurements employing an $AuSi_X$ intercalation layer between the $N_2HBC$-chains and the Au(111) substrate are shown in the Supplementary Information (Supplementary Note 5 and Supplementary Fig. S20).

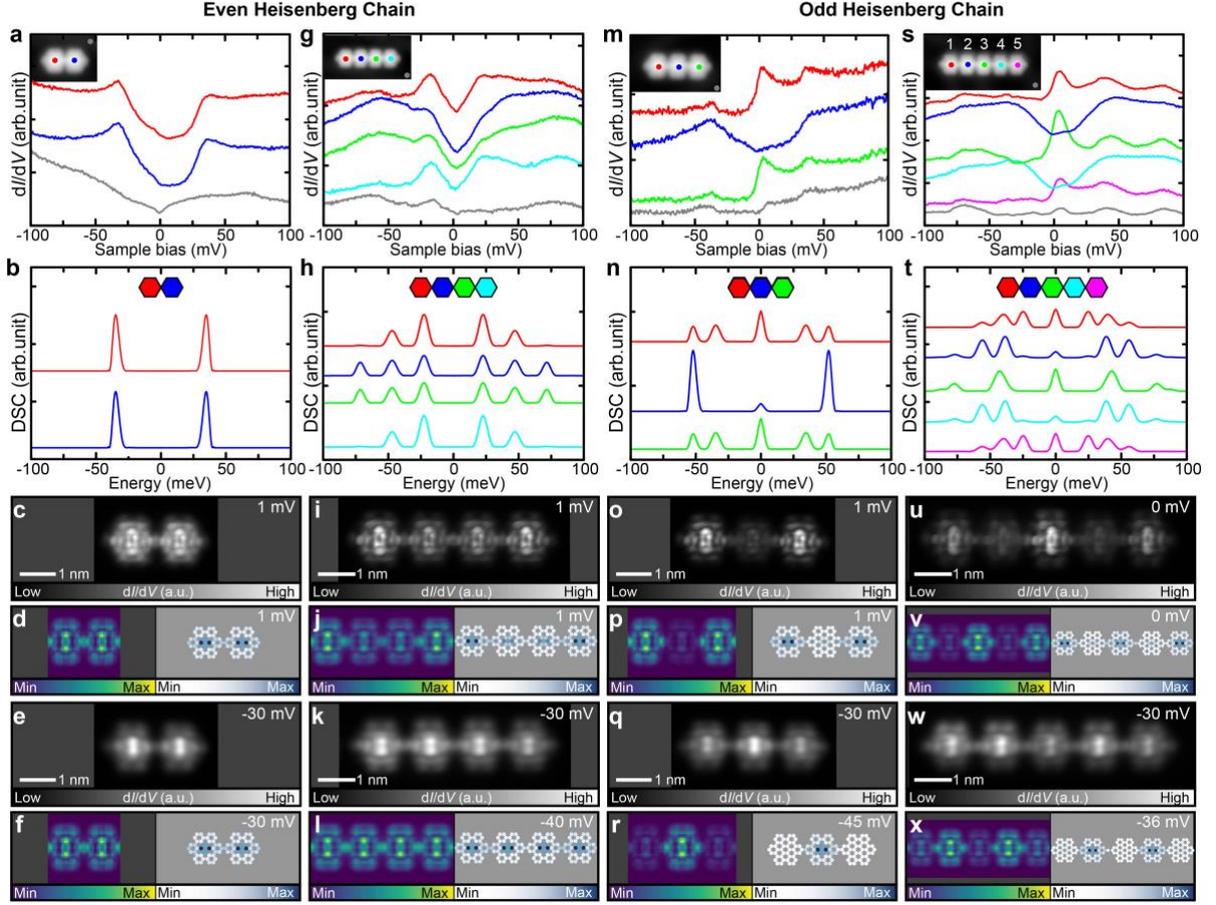

**Fig. 3. Magnetic properties of even and odd Heisenberg chains on Au(111).** Experimental d$I$/d$V$ curves and constant height maps and simulations for dimer **a-f**, tetramer **g-l**, trimer **m-r**, and pentamer **s-x**. The insets in (**a,g,m,s**) correspond to the STM topography of each chain, with the colored dots indicating the STS measurement positions (grey color is the signal from the Au(111) substrate for reference). The colored hexagons in (**b,h,n,t**) indicate in which unit of the chains the DSC was projected. **d,f,j,l,p,r,v,x**, Simulations were obtained by modulating the DFT calculated constant-height d$I$/d$V$ maps at 0 meV with the DSC at respective energies in the left panel and were acquired by modulating the spin densities calculated by mean-field Hubbard with the DSC in the right panel. Measurement parameters: $V$ = 1 mV and $V_{ac}$ = 10 mV in (**c,i,o**). $V$ = 0 mV and $V_{ac}$ = 1 mV in (**u**). $V$ = -30 mV, $I$ = 30 pA and $V_{ac}$ = 3 mV in (**e,k,q,w**).

**Conclusions**

We fabricated spin-1/2 antiferromagnetic molecular chains on Au(111) that depict different collective features depending on the parity of the chains: gapped excitations for an even number of repeating units or Kondo excitations for an odd number of repeating units. A combination of STM/STS experiments and DFT simulations were used to fully determine their chemical structures, as well as their electronic and magnetic properties, and the complex excitations observed in the low energy spectra were correctly addressed by a Heisenberg Hamiltonian with antiferromagnetic coupling between first neighbors. Our results provide a model realization of a quantum Heisenberg model, allowing to directly image the spatial distribution of its quantum many-body excitations. In particular, a molecular system with antiferromagnetic Heisenberg couplings presents a platform for the realization of topological spin models with fractionalized excitations (e.g. dimer chains and lattices) and frustrated spin systems.

**Methods**

STM experiments:

All experiments were performed in a low temperature scanning tunneling microscopy (STM) system (home-made) at 4.3 K under ultrahigh vacuum condition ($< 2 \times 10^{-10}$ mbar). A clean single crystal Au(111) substrate was prepared through cyclic sputtering ($Ar^+$, 10 min) and annealing (720 K, 15 min). The temperature of sample was measured by a thermocouple and a pyrometer. 2,13-dichloro-7b,7c,18b,18c-tetrahydro-3a$^2$,7c$^1$-diazahexabenzo[*a,cd,f,j,lm,o*]perylene (**1**) were deposited from Knudsen cells (Kentax GmbH). A STM tip was made from the chemically etched tungsten. For constant height d$I$/d$V$ imaging, the tip apex was functionalized by a CO molecule picked up from the surface. The bias voltage was set close to zero voltage. The modulation amplitude was 10 mV$_{0\text{-peak}}$ and the frequency was 510 Hz for the wide energy range STS measurements and bond resolved imaging. The modulation amplitude was 1 mV$_{0\text{-peak}}$ or 5 mV$_{0\text{-peak}}$ and the frequency was 510 Hz for the small energy range STS measurements.

Synthesis of compound **1**:

A solution of 2-chloro-8*H*-isoquinolino[4,3,2-*de*]phenanthridin-9-ium chloride (10 mg, 0.030 mmol) dissolved in dimethylsulfoxide (2.0 mL) was preheated to 190 °C for 30 seconds. Subsequently, tributylamine (0.17 mL, 0.71 mmol) was added and further stirred for 5 min. After cooling down to room temperature, the reaction mixture was extracted with dichloromethane. The combined organic phases were washed with water, dried under sodium sulfate, filtered, and evaporated *in vacuo*. The crude product was purified by silica gel column chromatography (hexane/dichloromethane = 8/1) to afford 2,13-dichloro-7b,7c,18b,18c-tetrahydro-3a$^2$,7c$^1$-diazahexabenzo[*a,cd,f,j,lm,o*]perylene (**1**) as a yellow solid (1.3 mg, 2 mmol, 7%). (See Supplementary Information for the details).

Theoretical calculations:

Spin-polarized DFT calculations were performed using the FHI-AIMS code[43]. Geometry optimizations were carried out twice, using two different exchange-correlation functionals: initial optimization with PBE[44], then followed by B3LYP[45], both with the standard 'light' basis set. For the gas phase modelled monomer, dimer, trimer, and the infinite chain, structural relaxations were only allowed in x-y plane (see the comparison of constrained and unconstrained relaxations in Supplementary Figure S14). For all the dI/dV simulations of monomer, the adsorption of monomer was first modelled on Au(111) with three atomic layers and two bottom layers were fully constrained. All free atoms were allowed to relax until the residual atomic forces reached less than 5e$^{-3}$ eV/Å. For the periodic chain, we used a distance of 15 Å to avoid interactions between the non-periodic images, and the geometry was relaxed using only the Γ-point. We used 8 times denser k-grid in the periodic direction to calculate band structures and density of states. The mean-field Hubbard calculations were realized with the PYQULA library[46] while the dynamical spin correlators were calculated using the DMRGPY libraty[41].

Theoretical constant height d*I*/d*V* maps were simulated using either fixed or relaxed CO tips, as specified in the relevant text. Simulations with fixed tips were computed by means of PPSTM code[47], employing a mixture of *s* and *p*$_{xy}$ waves - different ratios of these waves were considered (see Supplementary Information)[48,49]. For the simulations using relaxed tips, the PPAFM code was initially used to model the positions of the CO tips[50]. The lateral stiffness for the CO tip was set to 0.25 N/m and an oscillation amplitude of 1.0 Å was used. Subsequently the d$I$/d$V$ maps were generated by PPSTM with a mixed *sp*-wave tip in a ratio that best reproduces the experiments. We used a broadening parameter of 0.1 eV for all simulations to get a good simulation performance.

**ACKNOWLEDGEMENTS**

This work was supported in part by Japan Society for the Promotion of Science (JSPS) KAKENHI Grant Number 22H00285 for S.K., the Ministry of Education, Singapore, under its Academic Research Fund Tier 1 (RG2/23) for S.I., and the Academy of Finland (projects no. 346824) for N.C., O.S., P.L., Y.Y., Z.Z. and A.S.F. and (projects nos. 331342, 336243, and 349696) for A.S.F. and J.L.L. A.S.F. was supported by the World Premier International Research Center Initiative (WPI), MEXT, Japan. K.S. acknowledges the supporting of ICYS project. The authors acknowledge the computational resources provided by the Aalto Science-IT project and CSC, Helsinki.

# AUTHOR INFORMATION

## Contributions

S.K. conceived of the project. K.S. and S.K. performed the measurement and analyzed the data together with P.L. and A.S.F. N.C., O.S., A.O.F., J.L.L., and A.S.F performed the theoretical calculations. F.H. and S. I. synthesized the precursor molecules. K.S., O.S., A.S.F., and S.K. wrote the paper with input from all authors.

## Corresponding Authors

Correspondence to Shigeki Kawai, Adam S. Foster, or Shingo Ito.

*KAWAI.Shigeki@nims.go.jp

*adam.foster@aalto.fi

*sgito@ntu.edu.sg

## Ethics declarations

### Competing interests

The authors declare that they have no competing interests.

## Supporting Information

Supplementary Figs. 1–20, Supplementary Table 1, and Supplementary Note 1–5.

# Supplementary Materials for
# Heisenberg Spin-1/2 Antiferromagnetic Molecular Chains

Kewei Sun[1,2], Nan Cao[3], Orlando J. Silveira[3], Adolfo O. Fumega[3], Fiona Hanindita[4], Shingo Ito[4]\*, Jose L. Lado[3], Peter Liljeroth[3], Adam S. Foster[3,5]\*, Shigeki Kawai[2,6]\*

[1]*International Center for Young Scientists, National Institute for Materials Science, 1-2-1 Sengen, Tsukuba, Ibaraki 305-0047, Japan.*

[2]*Center for Basic Research on Materials, National Institute for Materials Sciences, 1-2-1 Sengen, Tsukuba, Ibaraki 305-0047, Japan.*

[3]*Department of Applied Physics, Aalto University, Espoo, Finland.*

[4]*Division of Chemistry and Biological Chemistry, School of Physical and Mathematical Sciences, Nanyang Technological University 21 Nanyang Link, 637371, Singapore.*

[5]*WPI Nano Life Science Institute (WPI-NanoLSI), Kanazawa University, Kakuma-machi, Japan.*

[6]*Graduate School of Pure and Applied Sciences, University of Tsukuba, Tsukuba 305-8571, Japan.*

**Synthesis of 1.**

**General**: All reactions were carried out under argon or nitrogen atmosphere using standard Schlenk techniques, unless otherwise noted. Thin-layer chromatography was performed using glass plates pre-coated with silica gel impregnated with a fluorescent indicator (Merck, #1.05715.0001). Silica gel column chromatography was performed as described by Still, et al.,[S1] employing silica gel 60N (spherical, neutral) purchased from W. R. Grace, Ltd.

**Instrumentation**: NMR spectra were recorded on Bruker Avance 400 and Bruker Avance III 400 spectrometers. Chemical shift values for protons are referenced to the signal of tetramethylsilane ($\delta = 0.00$) or the residual signal of dimethyl sulfoxide-$d_6$ ($\delta = 2.50$), and chemical shift values for carbons are referenced to the signal of tetramethylsilane ($\delta = 0.00$) or the carbon resonance of chloroform-$d$ ($\delta = 77.2$) and dimethyl sulfoxide-$d_6$ ($\delta = 39.5$). Infrared (IR) spectra were recorded on a PerkinElmer FTIR spectrum 100 with an attenuated total reflection (ATR) sampling accessary. High-resolution mass (HRMS) spectra were taken on a Waters Q-Tof Premier mass spectrometer with an electron spray ionization time-of-flight (ESI-TOF) method. Decomposition points and melting points were recorded on an OptiMelt MPA-100 apparatus.

**Materials**: The following reagents were purchased from indicated suppliers and used as received: benzo[*c*][1,2]oxaborol-1(3*H*)-ol (**II**; Fluorochem Ltd.), potassium carbonate ($K_2CO_3$, Tokyo Chemical Industry, Co. Ltd. (TCI)), tetrakis(triphenylphosphine)palladium(0) (Pd(PPh$_3$)$_4$; Sigma-Aldrich), tributylamine (Bu$_3$N, TCI), dimethylsulfoxide (DMSO; TCI). 4-Chloro-2,6-diiodoaniline was prepared according to a literature procedure[S2].

**Synthesis of Compound III**

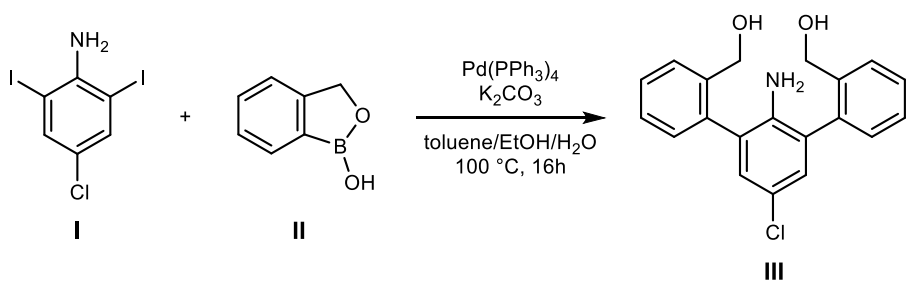

In a 25 mL Schlenk tube, 4-chloro-2,6-diiodoaniline (**I**, 0.20 g, 0.70 mmol), benzo[*c*][1,2]oxaborol-1(3*H*)-ol (**II**, 0.24 g, 1.75 mmol), tetrakis(triphenylphosphine)palladium(0) (Pd(PPh$_3$)$_4$, 41 mg, 0.035 mmol), and potassium carbonate (K$_2$CO$_3$, 0.78 g, 5.6 mmol) were dissolved in toluene (5.0 mL), ethanol (1.0 mL), and water (1.0 mL). After being stirred for 16 h at 100 °C, the mixture was allowed to cool down to room temperature, and extracted with ethyl acetate. The combined organic phase was washed with water, dried under sodium sulfate (Na$_2$SO$_4$), filtered, and evaporated to dryness. The crude product was purified by column chromatography (Hex: EA = 4:1 to 1:2) to afford compound **III** as a colorless solid (200 mg, 0.59 mmol, 84%).

Colorless solid; $R_f$ = 0.26 (hexane/ethyl acetate = 1/1); mp 193–195 °C; IR (neat) cm$^{-1}$ 3384, 3269, 1615, 1489, 1454, 1429, 1197, 1114, 1035, 1007, 950, 873, 764, 740, 710; In NMR spectra, two rotational isomers were observed in the ratio of *ca.* 6:4, $^1$H NMR (400 MHz, DMSO-*d*$_6$) *major* isomer: δ 7.63 (d, *J* = 7.5 Hz, 2H), 7.43 (t, *J* = 7.6 Hz, 2H), 7.35 (t, *J* = 7.0 Hz, 2H), 7.22–7.13 (m, 2H), 6.97 (s, 2H), 5.12 (t, *J* = 5.5 Hz, 2H), 4.45–4.28 (m, 4H), 3.76 (s, 2H), *minor* isomer: δ 7.62 (d, *J* = 7.2 Hz, 2H), 7.43 (t, *J* = 7.6 Hz, 2H), 7.36 (t, *J* = 7.4 Hz, 2H), 7.22–7.13 (m, 2H), 6.97 (s, 2H), 5.08 (t, *J* = 5.4 Hz, 2H), 4.45–4.28 (m, 4H), 3.77 (s, 2H); $^{13}$C NMR (101 MHz, DMSO-*d*$_6$) *major* isomer: δ 141.4, 141.3, 135.8, 130.0, 128.9, 128.5, 127.9, 127.6, 127.5, 120.3, 60.7. *minor* isomer: δ 141.4, 141.0, 136.0, 130.1, 128.9, 128.5, 127.9, 127.7, 127.6, 120.2, 60.9; HRMS (ESI) *m/z* calcd for C$_{20}$H$_{18}$ClNO$_2$ [M+H]$^+$ 340.1104, found 340.1115.

**Synthesis of Compound IV**

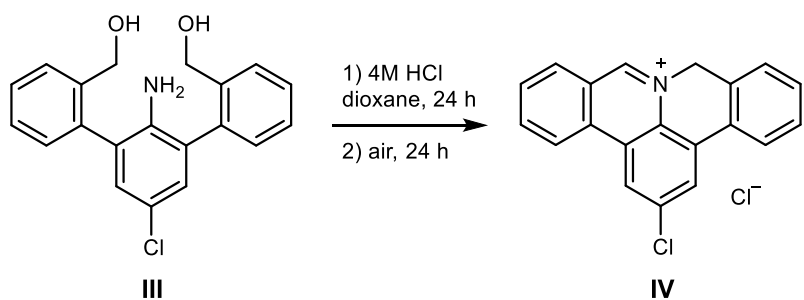

In a 50 mL Schlenk tube, compound **III** (0.2 mg, 0.59 mmol) was dissolved in 4M HCl in dioxane (12 mL, 48 mmol) under inert atmosphere. The mixture was stirred for 24 h at 120 °C.

After cooling down to room temperature, the cap of the Schlenk tube was removed. The mixture was stirred under air for 24 h at room temperature. The resulting precipitate was collected by filtration and washed with diethyl ether to afford compound **IV** as a yellow solid (186 mg, 0.55 mmol, 93%).

Yellow solid; mp 225–227 (dec.); IR (neat) cm$^{-1}$ 3066, 1622, 1596, 1579, 1531, 1504, 1423, 1392, 1348, 1244, 1230, 1133, 1119, 1047, 903, 872, 839, 786, 756, 728, 708; $^1$H NMR (400 MHz, DMSO-$d_6$) δ 10.38 (s, 1H), 9.18 (s, 1H), 9.17 (d, $J$ = 5.9 Hz, 1H), 8.80 (d, $J$ = 1.6 Hz, 1H), 8.66 (t, $J$ = 7.7 Hz, 1H), 8.44–8.33 (m, 2H), 8.15 (t, $J$ = 7.5 Hz, 1H), 7.64–7.55 (m, 3H), 6.28 (s, 2H); $^{13}$C NMR (101 MHz, DMSO-$d_6$) δ 155.0, 138.6, 136.8, 133.6, 133.1, 131.7, 131.1, 129.8, 129.1, 129.1, 129.0, 128.1, 126.9, 126.4, 125.9, 125.1, 124.7, 124.4, 123.8, 57.5; HRMS (ESI) $m/z$ calcd for $C_{20}H_{13}ClN$ [M]$^+$ 302.0737, found 302.0739.

**Synthesis of Compound 1**

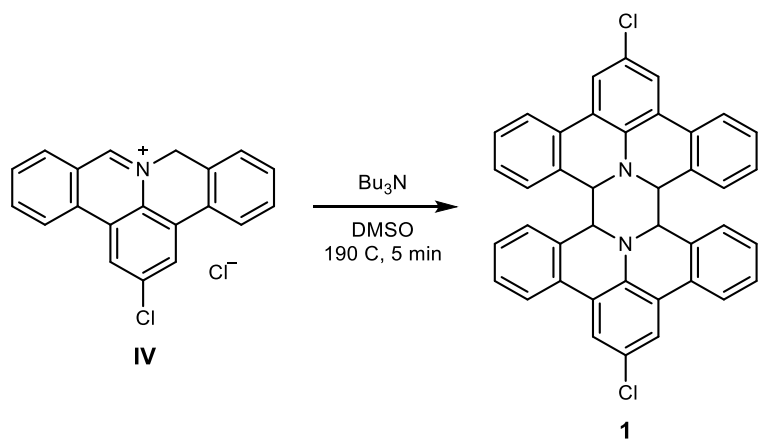

A solution of compound **IV** (10 mg, 0.030 mmol) dissolved in diemthylsulfoxide (DMSO; 2.0 mL) was preheated to 190 °C for 30 seconds. Subsequently, tributylamine (0.17 mL, 0.71 mmol) was added and further stirred for 5 min. After cooling down to room temperature, the reaction mixture was extracted with dichloromethane. The combined organic phases were washed with water, dried under sodium sulfate (Na$_2$SO$_4$), filtered, and evaporated *in vacuo*. The crude product was purified by silica gel column chromatography (hexane/dichloromethane = 8/1) to afford compound **1** as a yellow solid (1.3 mg, 2 mmol, 7%).

Yellow solid; $R_f$ = 0.30 (hexane/dichloromethane = 4/1); mp >300 °C; IR (neat) cm$^{-1}$ 2921, 2852, 1708, 1577, 1496, 1481, 1432, 1388, 1327, 1297, 1229, 1160, 1104, 1049, 941, 841, 778,

752, 730, 686; $^1$H NMR (400 MHz, CDCl$_3$) δ 7.71 (s, 4H), 7.59 (d, *J* = 8.0 Hz, 4H), 7.26 (t, *J* = 8.0 Hz, 4H), 6.86 (t, *J* = 8.0 Hz, 4H), 6.02 (d, *J* = 8.0 Hz, 4H), 5.11 (s, 4H); $^{13}$C NMR (101 MHz, CDCl$_3$) δ 139.1 (2C), 129.7 (4C), 129.09 (4C), 128.8 (4C), 128.5 (4C), 127.3 (4C), 124.5 (2C), 124.2 (4C), 123.1 (4C), 122.0 (4C), 61.5 (4C); HRMS (ESI) *m/z* calcd for C$_{40}$H$_{24}$Cl$_2$N$_2$ [M+H]$^+$ 603.1395, found 603.1399.

**Supplementary Note 1**

**DFT simulations of STM and d$I$/d$V$ maps:**

Fig. S5a,b show two experimental d$I$/d$V$ maps of the N$_2$HBC$^+$ molecule acquired at zero bias. The simulations in Fig. S5c calculated with a flexible CO tip show that the features of Fig. S5a are correctly captured by the simulations at near tip-sample distance, especially at the brighter inner part of the molecule. In this scenario, the flexibility of the CO tip is important to enhance the contrast of the image. As the tip-sample distance increases, the flexibility of the CO tip becomes less important, and the simulations start to resemble more the experimental image in Fig. S5b. Figure S6 shows a series of d$I$/d$V$ maps acquired with a rigid CO tip, resembling even more the experimental image in Fig. S5b. Figure S6 also shows simulated d$I$/d$V$ maps calculated with different tip conditions, where we found that a linear combination of 5% of *s* and 95% of *p$_{xy}$* of the tip orbitals best resembles the experimental images.

**Supplementary Note 2**

**Mean-field Hubbard *vs.* spin-polarized DFT calculations of periodic chains:**

We investigated the electronic and magnetic properties of the periodic $N_2HBC^+$ chains by using mean-field Hubbard model and spin-polarized DFT, which have been widely used to determine the magnetization of nanographenes and demonstrate agreement with experiments[S3]. Both mean-field Hubbard and spin-polarized DFT calculations predict the antiferromagnetic alignment to be energetically favored (details can be found in the main text). The presence of a nitrogen is included in the tight binding model as a site with a local negative onsite energy, and its different charge state is accounted for in the filling of the system. We performed self-consistent calculations of the tight binding model, allowing access to the self-consistent magnetization, charge density and electronic structure[S4]. In the absence of spin polarization, the periodic chains display metallic band structures, as shown in Fig. S12a,b and S13b. Spin polarization results in an increased bandgap and the localization of electronic states at the center of each unit with an antiferromagnetic ordering, as shown in the calculated spin density maps (Fig. S12c,d and S13a,c). The calculated spin density distributions align with the experimentally resolved spin localization at the center of each unit of the chains (Figure 3 of the main text). In our calculation, we use only the nearest neighbors hopping. The hopping term *t* is set to a constant value of 2.7 eV. As shown in Fig. S12e, the mean-field Hubbard model calculations show that the exchange interaction *J* for periodic chains varies substantially as a function of the on-site Coulomb repulsion *U*. This variation suggests the sensitive dependence of the exchange interaction on the on-site electronic interaction characterized by *U*. We find that exchange couplings compatible with the DFT calculations and our experimental observation are obtained for a value of $U = 1.5\ t$, similar to other nanographene systems.

**Supplementary Note 3**

**DFT calculations of the antiferromagnetic chains:**

PBE/DFT-calculated work functions of neutral and positively charged chains (Table S1) shows that charge transfer is always necessary to allow the Fermi level alignment of the chains with gold in relation to the vacuum. For the $N_2HBC$ molecule, dimer- and trimer-$N_2HBC$, our calculations clearly indicate that all systems will be positively charged due to a local transfer of approximately one electron to gold from each $N_2HBC$ unit. For larger chains (tetramer and pentamer), charging the whole system in the calculation led to unphysical results due to delocalized effects of the removal of the electrons that are not consistent with the local charge transfer between system and substrate. Nevertheless, we conclude that each $N_2HBC$ unit donates approximately one electron to the Au(111) substrate regardless of the size of the chain. We also propose that the same behavior will be observed for even longer chains, approaching the infinite-chain regime.

Regarding the geometrical structure of the $N_2HBC$ molecule and chains, we noticed that they are strongly distorted if fully relaxed (Fig. S15), indicating that, besides the charge transfer, the Au(111) substrate is necessary for structural stability of the chains. Thus, all the free-standing calculations were realized allowing relaxation only in the x-y plane.

After geometry relaxation, the energy levels and molecular orbitals shown in Fig. S16 and S17 were obtained within the B3LYP level. In Fig. S16a, the SOMO-SUMO gap of the $N_2HBC^+$ molecule is 1.26 eV, and both SOMO and SUMO in Fig. S16b have the same spatial distributions. The SOMO-SUMO gap of the open shell singlet dimer-$N_2HBC^{2+}$ in Fig. S16c is 1.22 eV, and the SOMO and SUMO have opposite spatial distributions depending on the spin (Fig. S16d). On the other hand, Fig. S16e shows that the SOMO-SUMO gap of the triplet dimer-$N_2HBC^{2+}$ is much smaller (0.99 eV), which also changes completely the spatial distribution of the orbitals (Fig. S16f). The triplet dimer-$N_2HBC^{2+}$ is 27 meV higher in energy than its open-shell singlet ground state, which is comparable to the experimental value of the super exchange coupling shown in Figure S10. The energy levels and molecular orbitals of the trimer-$N_2HBC^{3+}$ are shown in Fig. S17. The low-energy molecular orbitals of the trimer-$N_2HBC^{3+}$ in the $S = 1/2$ and $S = 3/2$ states are either spatially very localized at its central $N_2HBC^+$ unit or at its peripheral units. The S = 3/2 trimer-$N_2HBC^{3+}$ is 62 meV higher in energy than its S = 1/2 ground state.

**Supplementary Note 4**

**Analytical solution for three interacting spin ½ electrons in a linear chain**

The Heisenberg Hamiltonian of such system can be written as:

$$\hat{H} = J(\vec{S_1} \cdot \vec{S_2}) + J(\vec{S_2} \cdot \vec{S_3})$$

which can be expressed in terms of the spin-ladder operators $S_n^{\pm} = S_n^x \pm i S_n^y$:

$$\vec{S_1} \cdot \vec{S_2} = S_1^x S_2^x + S_1^y S_2^y + S_1^z S_2^z = \frac{1}{2}(S_1^+ S_2^- + S_1^- S_2^+) + S_1^z S_2^z.$$

For three spins, the total spin $S_T$ can take values of a doublet $\pm 1/2$ and quartet $\pm 3/2$. The $2^3$ possible states for the three spin 1/2 system are:

$$|\uparrow\uparrow\uparrow\rangle, |\uparrow\uparrow\downarrow\rangle, |\uparrow\downarrow\uparrow\rangle, |\uparrow\downarrow\downarrow\rangle, |\downarrow\uparrow\uparrow\rangle, |\downarrow\uparrow\downarrow\rangle, |\downarrow\downarrow\uparrow\rangle, |\downarrow\downarrow\downarrow\rangle.$$

Hence, taking $J$ as a common factor, the Hamiltonian matrix takes the following form in this basis:

|   | $\|\uparrow\uparrow\uparrow\rangle$ | $\|\uparrow\uparrow\downarrow\rangle$ | $\|\uparrow\downarrow\uparrow\rangle$ | $\|\uparrow\downarrow\downarrow\rangle$ | $\|\downarrow\uparrow\uparrow\rangle$ | $\|\downarrow\uparrow\downarrow\rangle$ | $\|\downarrow\downarrow\uparrow\rangle$ | $\|\downarrow\downarrow\downarrow\rangle$ |
|---|---|---|---|---|---|---|---|---|
| $\langle\uparrow\uparrow\uparrow\|$ | ½ | 0 | 0 | 0 | 0 | 0 | 0 | 0 |
| $\langle\uparrow\uparrow\downarrow\|$ | 0 | 0 | ½ | 0 | 0 | 0 | 0 | 0 |
| $\langle\uparrow\downarrow\uparrow\|$ | 0 | ½ | -½ | 0 | ½ | 0 | 0 | 0 |
| $\langle\uparrow\downarrow\downarrow\|$ | 0 | 0 | 0 | 0 | 0 | ½ | 0 | 0 |
| $\langle\downarrow\uparrow\uparrow\|$ | 0 | 0 | ½ | 0 | 0 | 0 | 0 | 0 |
| $\langle\downarrow\uparrow\downarrow\|$ | 0 | 0 | 0 | ½ | 0 | -½ | ½ | 0 |
| $\langle\downarrow\downarrow\uparrow\|$ | 0 | 0 | 0 | 0 | 0 | ½ | 0 | 0 |
| $\langle\downarrow\downarrow\downarrow\|$ | 0 | 0 | 0 | 0 | 0 | 0 | 0 | ½ |

The two many-body doublet ground states can be written as:

$$|\alpha\rangle = \frac{1}{\sqrt{6}}(|\uparrow\uparrow\downarrow\rangle - 2|\uparrow\downarrow\uparrow\rangle + |\downarrow\uparrow\uparrow\rangle) \text{ and } |\beta\rangle = \frac{1}{\sqrt{6}}(|\uparrow\downarrow\downarrow\rangle - 2|\downarrow\uparrow\downarrow\rangle + |\uparrow\downarrow\downarrow\rangle),$$

which are the states that contribute to the spin correlator at zero bias. The other excited many-body states are:

$$|\varphi_1\rangle = \frac{1}{\sqrt{2}}(|\uparrow\uparrow\downarrow\rangle - |\downarrow\uparrow\uparrow\rangle); \qquad |\varphi_2\rangle = \frac{1}{\sqrt{2}}(|\uparrow\downarrow\downarrow\rangle - |\uparrow\downarrow\downarrow\rangle);$$

$$|\varphi_3\rangle = \frac{1}{\sqrt{3}}(|\uparrow\uparrow\downarrow\rangle + |\uparrow\downarrow\uparrow\rangle + |\downarrow\uparrow\uparrow\rangle); \qquad |\varphi_4\rangle = \frac{1}{\sqrt{3}}(|\uparrow\downarrow\downarrow\rangle + |\downarrow\uparrow\downarrow\rangle + |\downarrow\downarrow\uparrow\rangle);$$

$|\varphi_5\rangle = |\uparrow\uparrow\uparrow\rangle;$   $\quad |\varphi_6\rangle = |\downarrow\downarrow\downarrow\rangle$

**Supplementary Note 5**

**Magnetic properties of $N_2$HBC chains on a AuSi$_X$ layer formed on Au(111)**

To verify the robustness of magnetic properties of the $N_2$HBC chains, we formed a AuSi$_X$ intercalation layer between the nanostructures and the Au(111) substrate by depositing Si atoms on the sample[S5-S7]. The STM topography shows $N_2$HBC chains and GNR 4 oligomers on the AuSi$_X$/Au(111) surface (Fig. S20a). A close-up image exhibits individual monomer **3**, in which the morphology is approximately symmetrical (inset of Fig. S20b). A Kondo resonance zero-bias peak is detected in the d$I$/d$V$ spectrum (Fig. S20b) recorded over the $N_2$HBC molecule, demonstrating a stronger intensity at the molecular center, as shown in the d$I$/d$V$ map at 0 mV (Fig. S20c). In contrast, an individual dimer-$N_2$HBC, reveals a slight asymmetric signal probably due to the underlying heterogeneous AuSi$_X$ area (inset of Fig. S20d). d$I$/d$V$ curves (Fig. S20d) recorded over the dimer-$N_2$HBC exhibit unprecedented spin excitation spectra, in which two sharp peaks at ±43 mV are unambiguously identified. Spatial distribution of spin excitation states is shown in d$I$/d$V$ maps at ±43 mV in Fig. S20e,f, revealing approximately symmetrical patterns. Trimer-$N_2$HBC (inset of Fig. S20g) is also slightly asymmetrical as the case of dimer-$N_2$HBC. d$I$/d$V$ curves (Fig. S20g) recorded over the trimer-$N_2$HBC show the zero-bias peaks at the unit on both sides, yet step-like features at the central unit, which is similar to observations on Au(111). Fig. S20h,i exhibit the d$I$/d$V$ maps at 0 mV and 40 mV over the-trimer-$N_2$HBC, revealing the spatial distribution of the Kondo resonance state and spin excitation state.

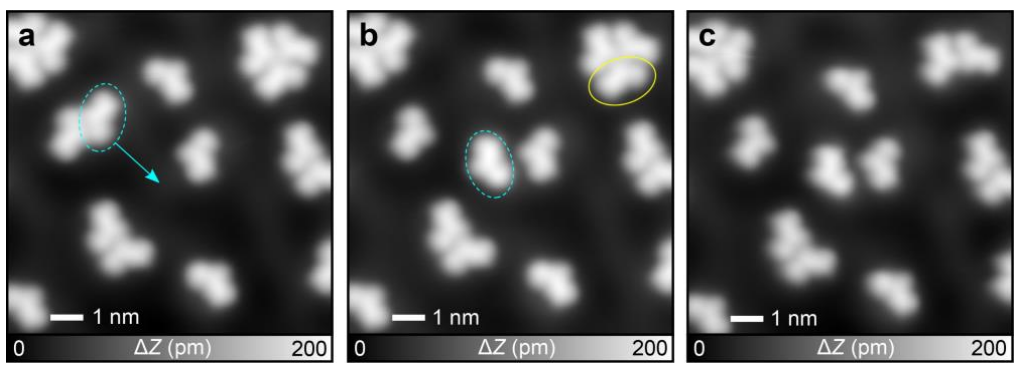

**Fig. S1. A series of tip-induced manipulations of molecules 2. a**, STM topography of as-deposited **1** on Au(111). The blue ellipse and arrow indicate the target molecule **2** and the trajectory of tip movement. **b**, STM topography taken after tip-induced manipulation. The blue ellipse indicates the manipulated molecule **2**. The yellow ellipse indicates the tip approaching another **2**. **c**, STM topography taken after the molecule **2** adsorbed onto the tip apex. Measurement parameters: $V = 50$ mV and $I = 10$ pA.

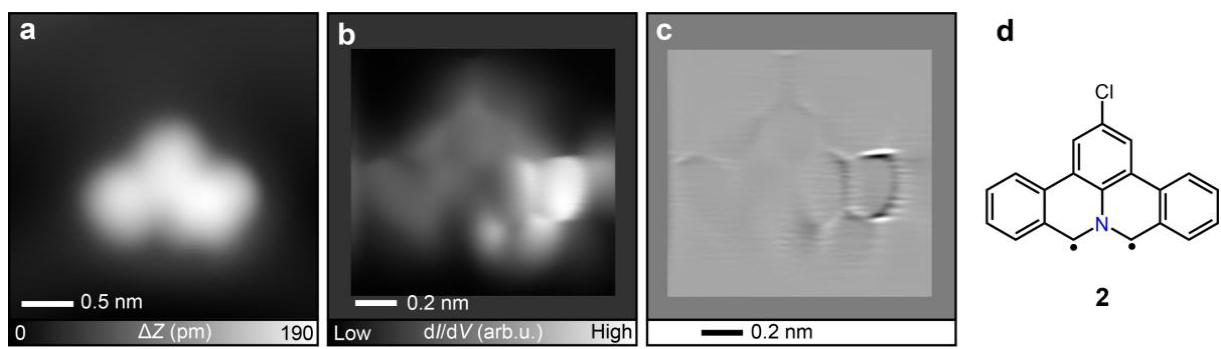

**Fig. S2. a**, STM topography of an individual **2** on Au(111). **b-d**, Constant height d$I$/d$V$ map (**b**), the corresponding Laplace filtered image (**c**), and Chemical structure of **2** (**d**). Measurement parameters: $V = 50$ mV and $I = 5$ pA.

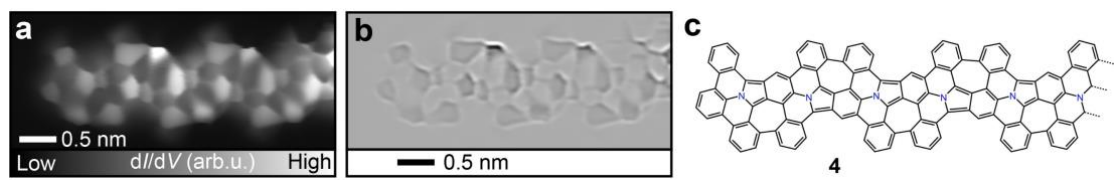

**Fig. S3**. Bond-resolved STM image of individual **4** on Au(111) in **a**, the corresponding Laplace filtered image in **b**, and the chemical structure in **c**. Measurement parameters: $V$ = 1 mV and $V_{ac}$ = 10 mV in (**a**).

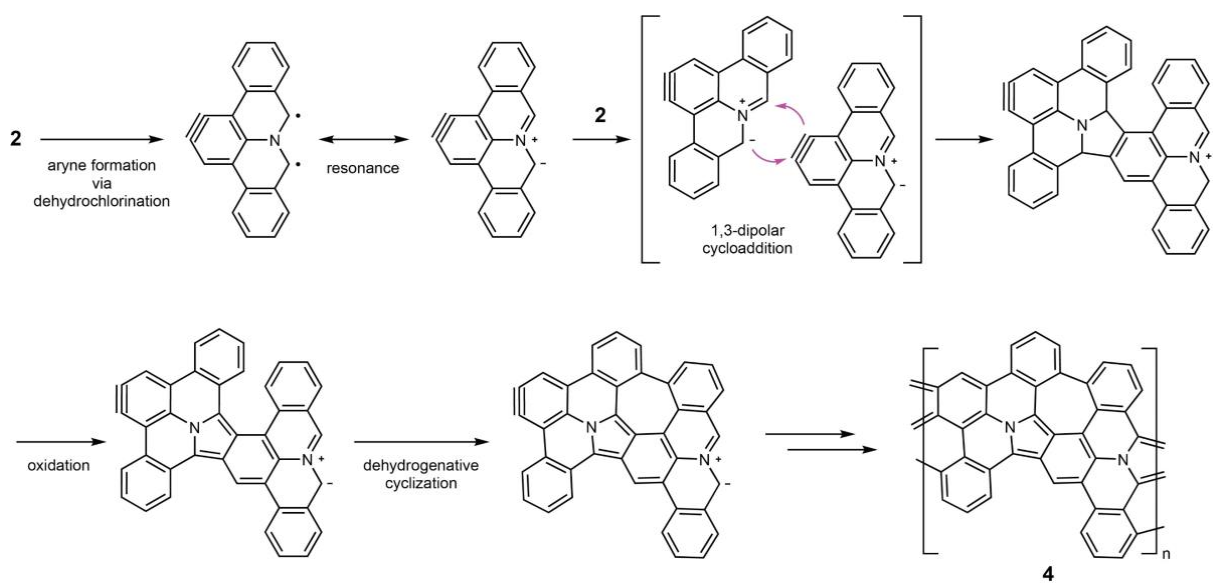

**Fig. S4**. Reaction path to synthesize **4**.

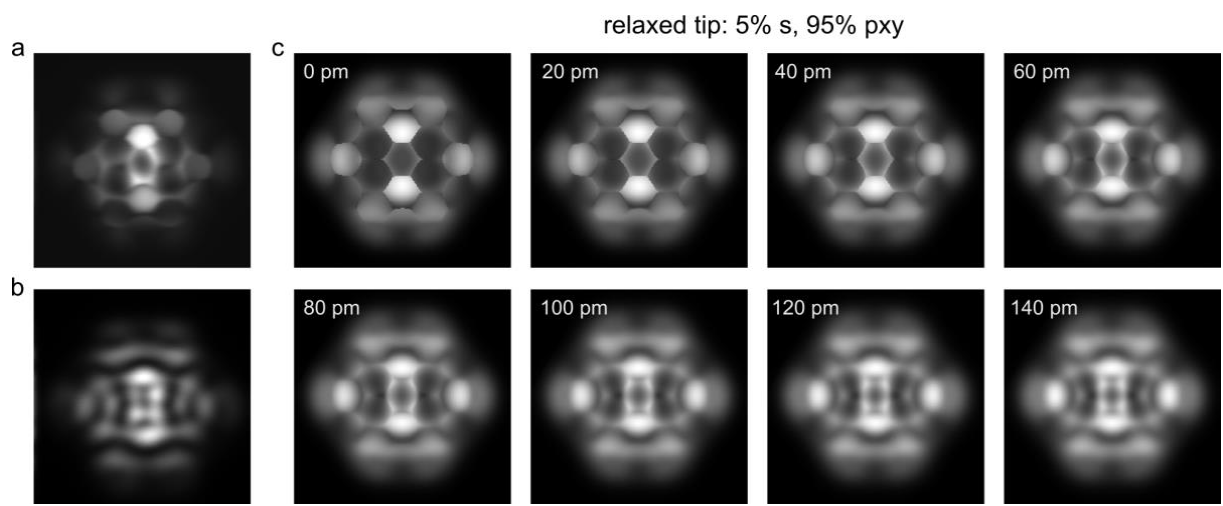

**Fig. S5. a,b,** Two different experimental d*I*/d*V* maps acquired at zero bias voltage at different tip-sample distances. **c**, Constant height d*I*/d*V* simulations were realized considering a flexible CO tip.

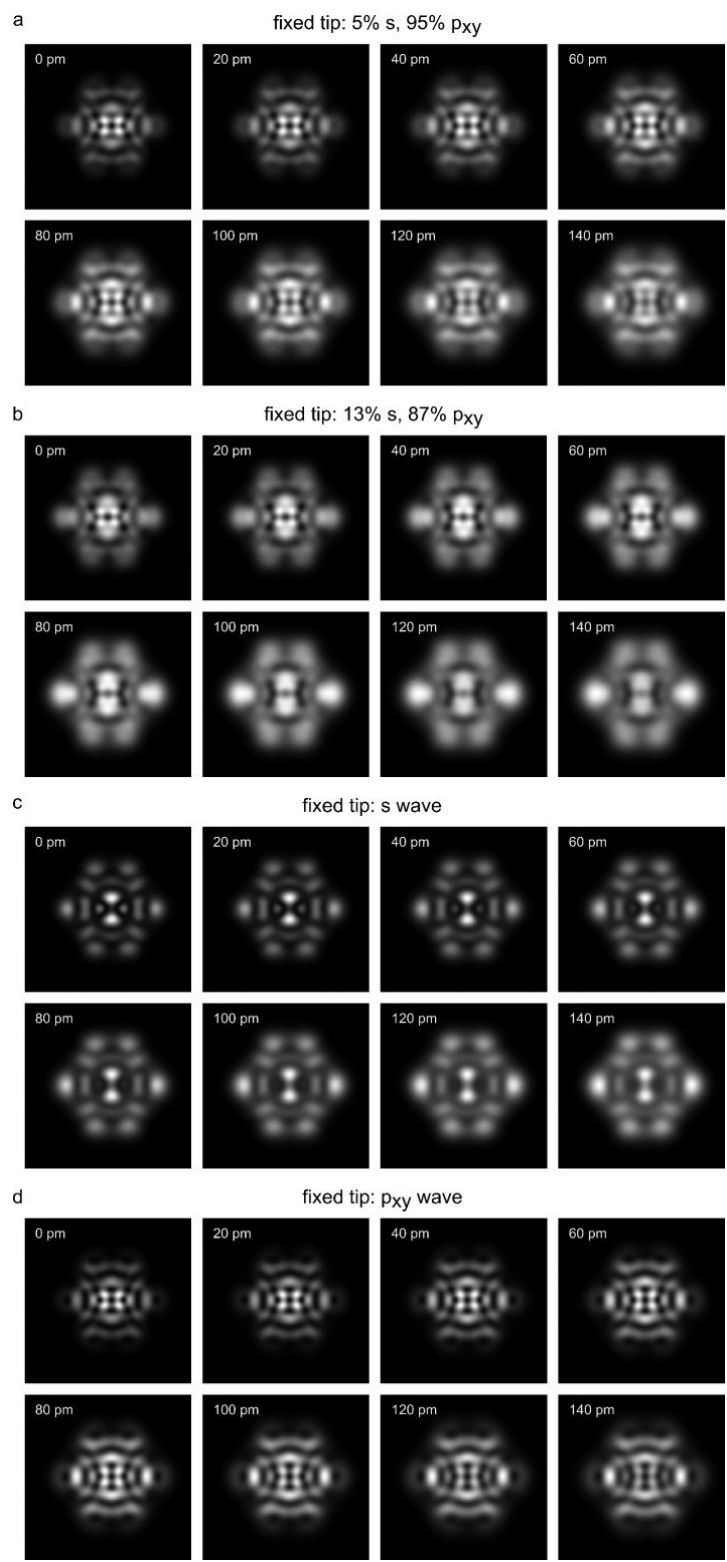

**Fig. S6.** A series of constant height d$I$/d$V$ simulations obtained with mixed $s$ and $p_{xy}$-wave tips, which were acquired through a linear combination of $s$ and $p_{xy}$ orbitals with different weights. The mixing was realized to simulate the effect of a rigid CO tip attached to the tip apex.

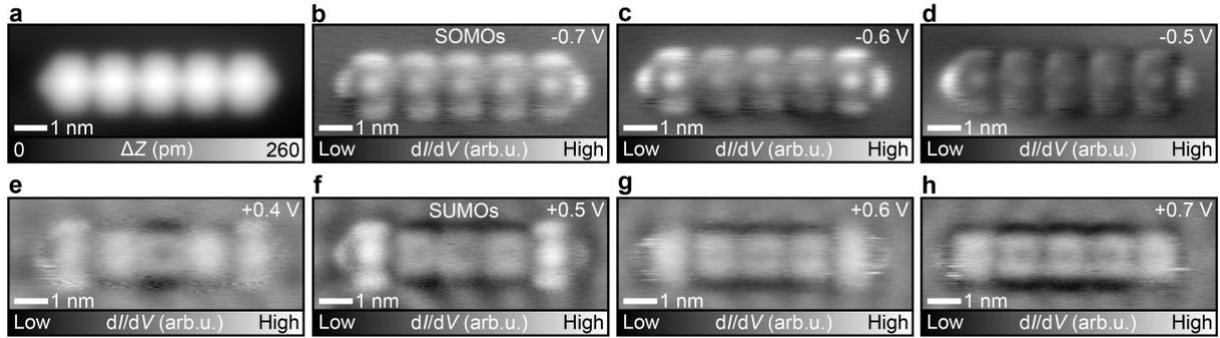

**Fig. S7. Electronic properties of an individual pentamer of N$_2$HBC. a,** STM topography. **b-h**, Constant current d$I$/d$V$ maps measured at different bias voltages around the energy of SUMOs and SUMOs. Measurement parameters: $V$ = 200 mV and $I$ = 10 pA in (**a**). For STS measurement, the tip-sample distance was adjusted at the corresponding bias voltages and $I$ = 160 pA, $V_{ac}$ = 10 mV in (**b-d**) (**f-h**). $I$ = 150 pA, $V_{ac}$ = 10 mV in (**e**).

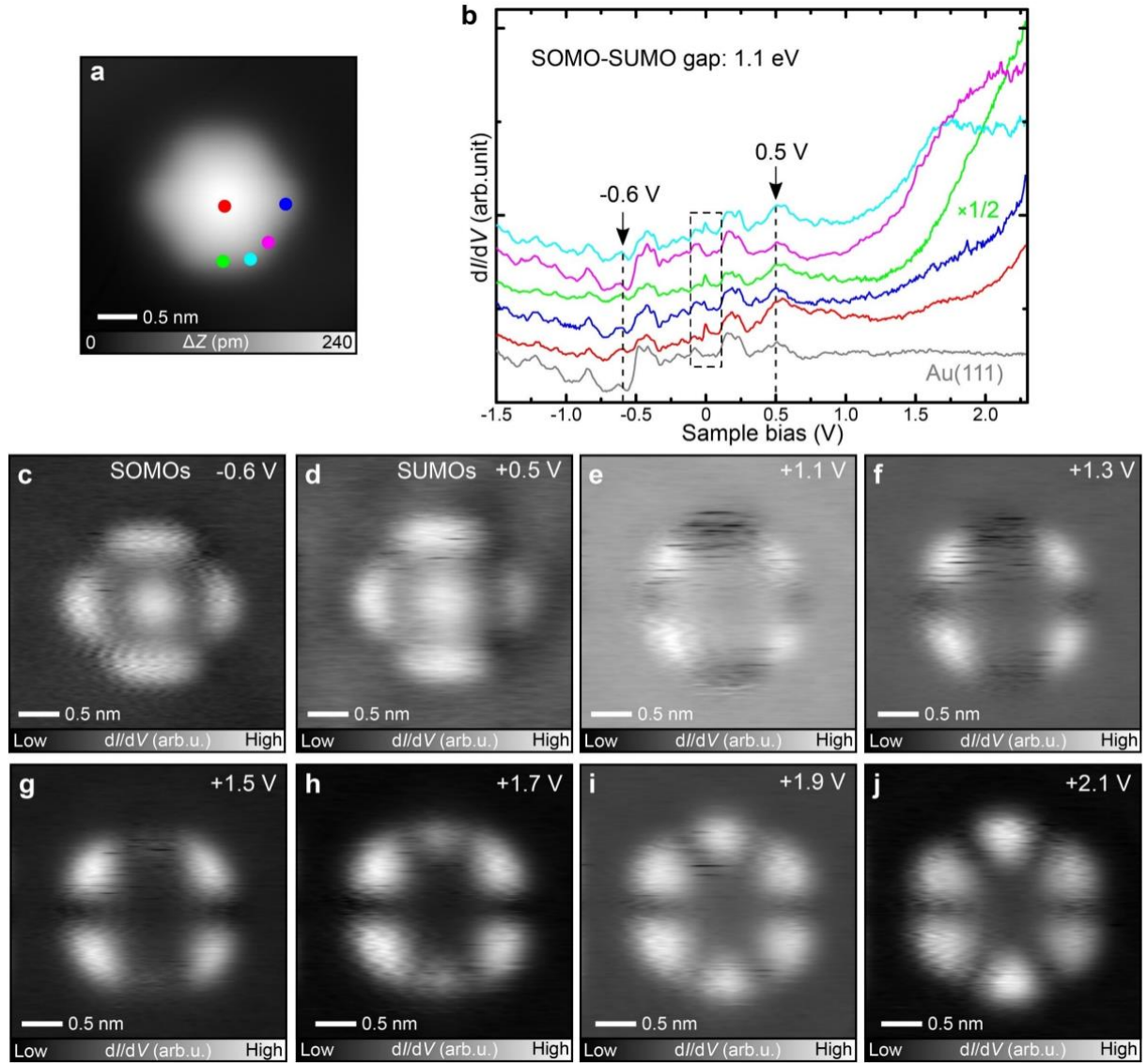

**Fig. S8. Electronic properties of an individual 3**. **a**, STM topography. **b**, d$I$/d$V$ curves recorded at different sites as indicated by colored dots in (**a**). **c-j**, Constant current d$I$/d$V$ maps measured at different bias voltages. Measurement parameters: $V$ = 200 mV and $I$ = 10 pA on (**a**).

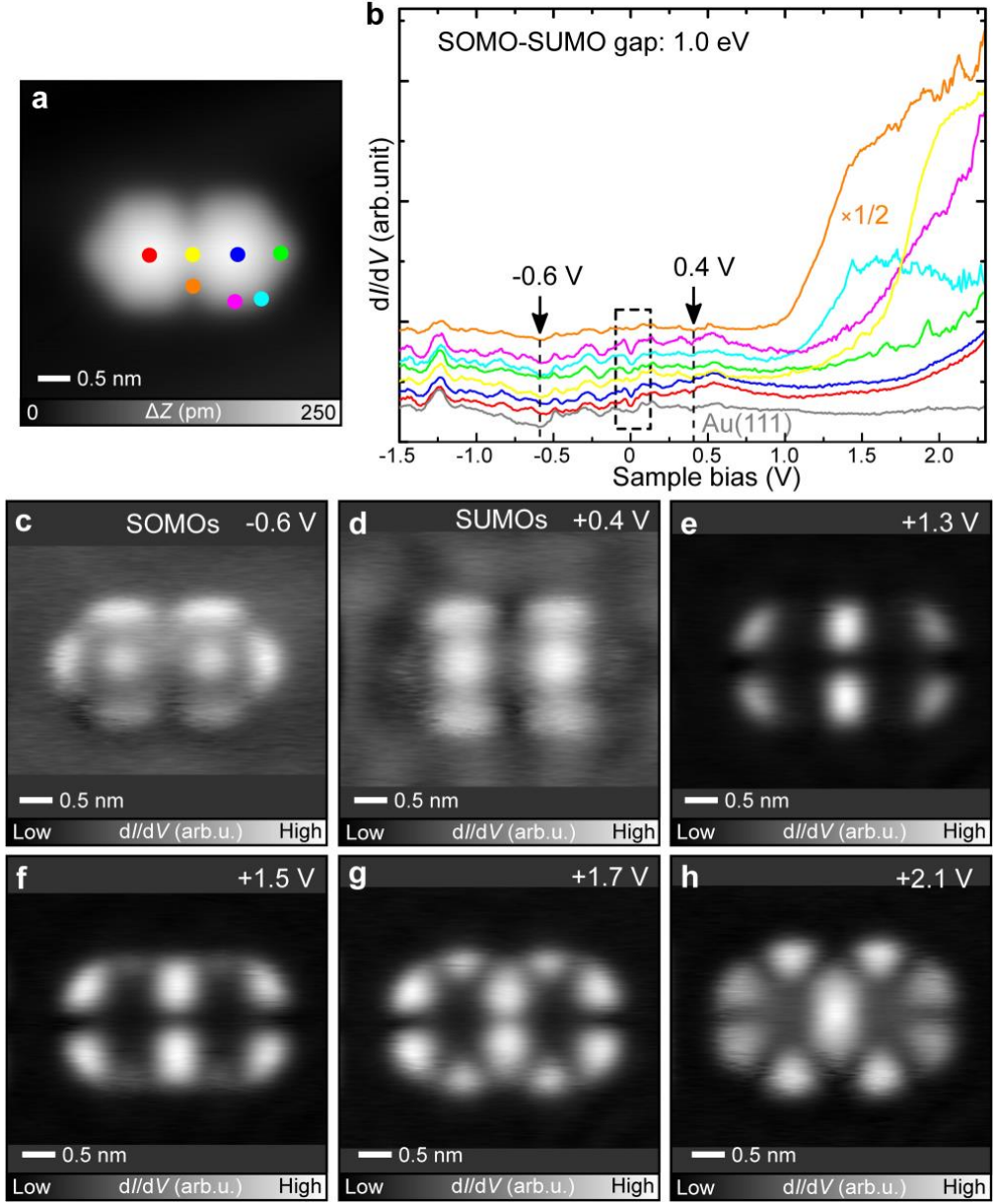

**Fig. S9. Electronic properties of a dimer of N$_2$HBC. a,** STM topography. **b,** d$I$/d$V$ curves recorded at different sites as indicated by colored dots in (**a**). **c-h**, Constant current d$I$/d$V$ maps measured at different bias voltages. Measurement parameters: $V$ = 200 mV and $I$ = 10 pA in (**a**).

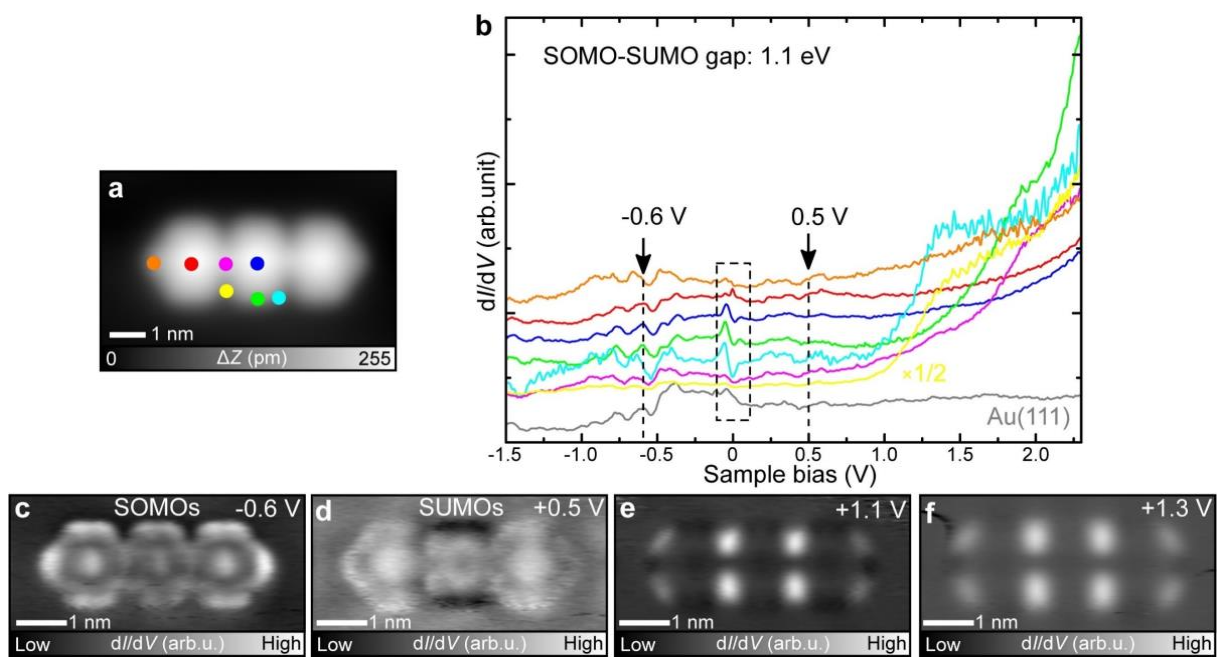

**Fig. S10. Electronic properties of a trimer of N$_2$HBC. a**, STM topography. **b**, d$I$/d$V$ curves recorded at different sites as indicated by colored dots in (**a**). **c-f**, Constant current d$I$/d$V$ maps measured at different bias voltages. Measurement parameters: $V$ = 200 mV and $I$ = 10 pA in (**a**).

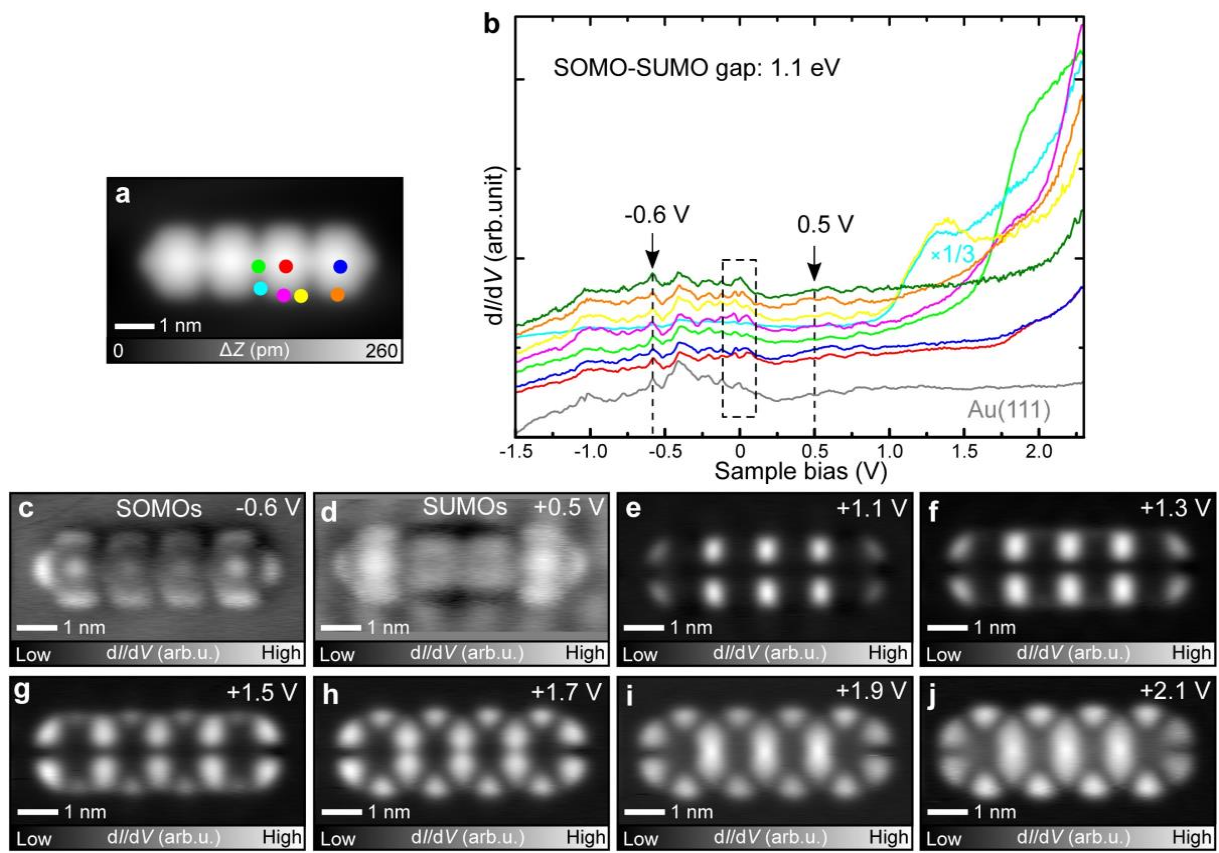

**Fig. S11. Electronic properties of a tetramer of N$_2$HBC. a**, STM topography. **b**, d$I$/d$V$ curves recorded at different sites as indicated by colored dots in (**a**). **c-j**, Constant current d$I$/d$V$ maps measured at different bias voltages. Measurement parameters: $V$ = 200 mV and $I$ = 10 pA in (**a**).

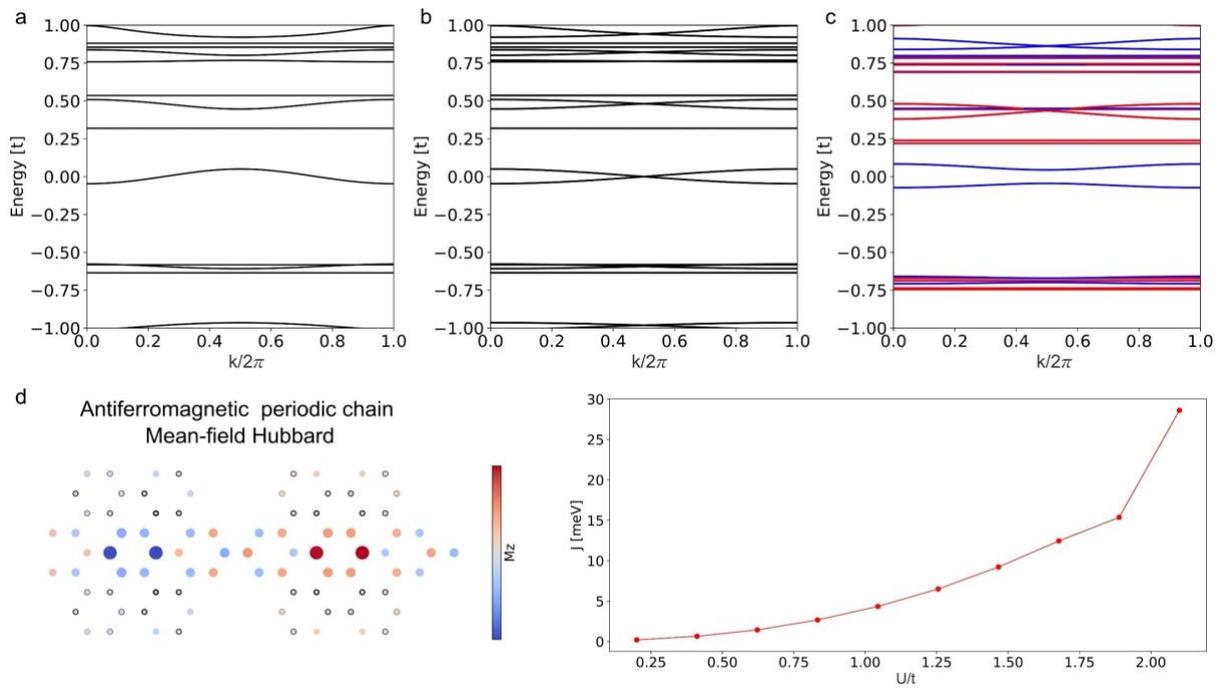

**Fig. S12.** Mean-field Hubbard calculations of periodic $N_2HBC^+$ chains. Band structures of the periodic mono-$N_2HBC^+$ chain **a**, periodic dimer-$N_2HBC^{2+}$ chain **b**, and the periodic dimer-$N_2HBC^{2+}$ chain with an antiferromagnetic configuration **c**. **d**, Spin density distributions of periodic antiferromagnetic chains from mean-field Hubbard simulations (on-site Coulomb parameter was set to 2.0 eV). Red, spin up; Blue, spin down. **e**, Exchange coupling energies $J$ between the excitation state and antiferromagnetic ground state of the periodic chain plotted as a function of $U/t$. The hopping parameter $t$ between the nearest neighbors is 2.7 eV.

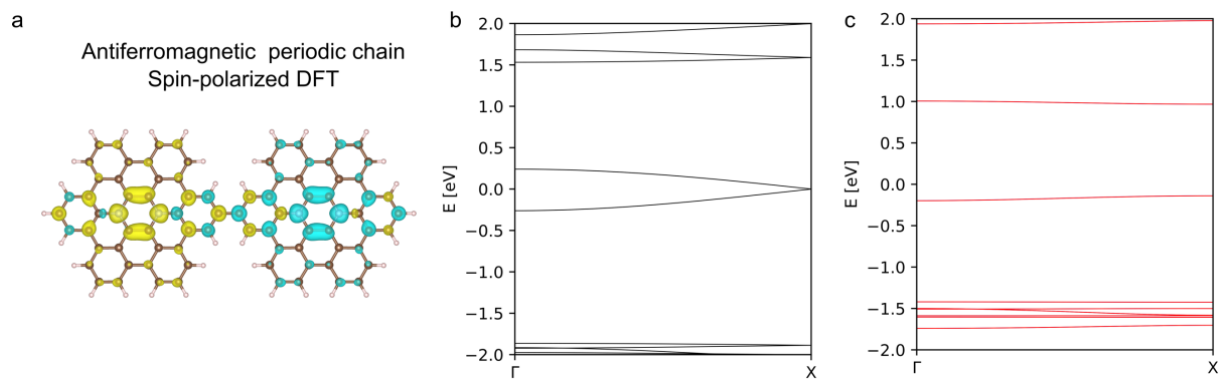

**Fig. S13. Spin-polarized DFT calculations of periodic N$_2$HBC$^+$ chains. a**, Spin density distributions of periodic antiferromagnetic chains. Green, spin up; yellow, spin down. **b**, Band structure of periodic dimer-N$_2$HBC$^{2+}$ chains without considering spin polarization. **c**, Band structure of spin-polarized periodic dimer-N$_2$HBC$^{2+}$ chains with antiferromagnetic configuration (duplicate from Figure 2h in the main text).

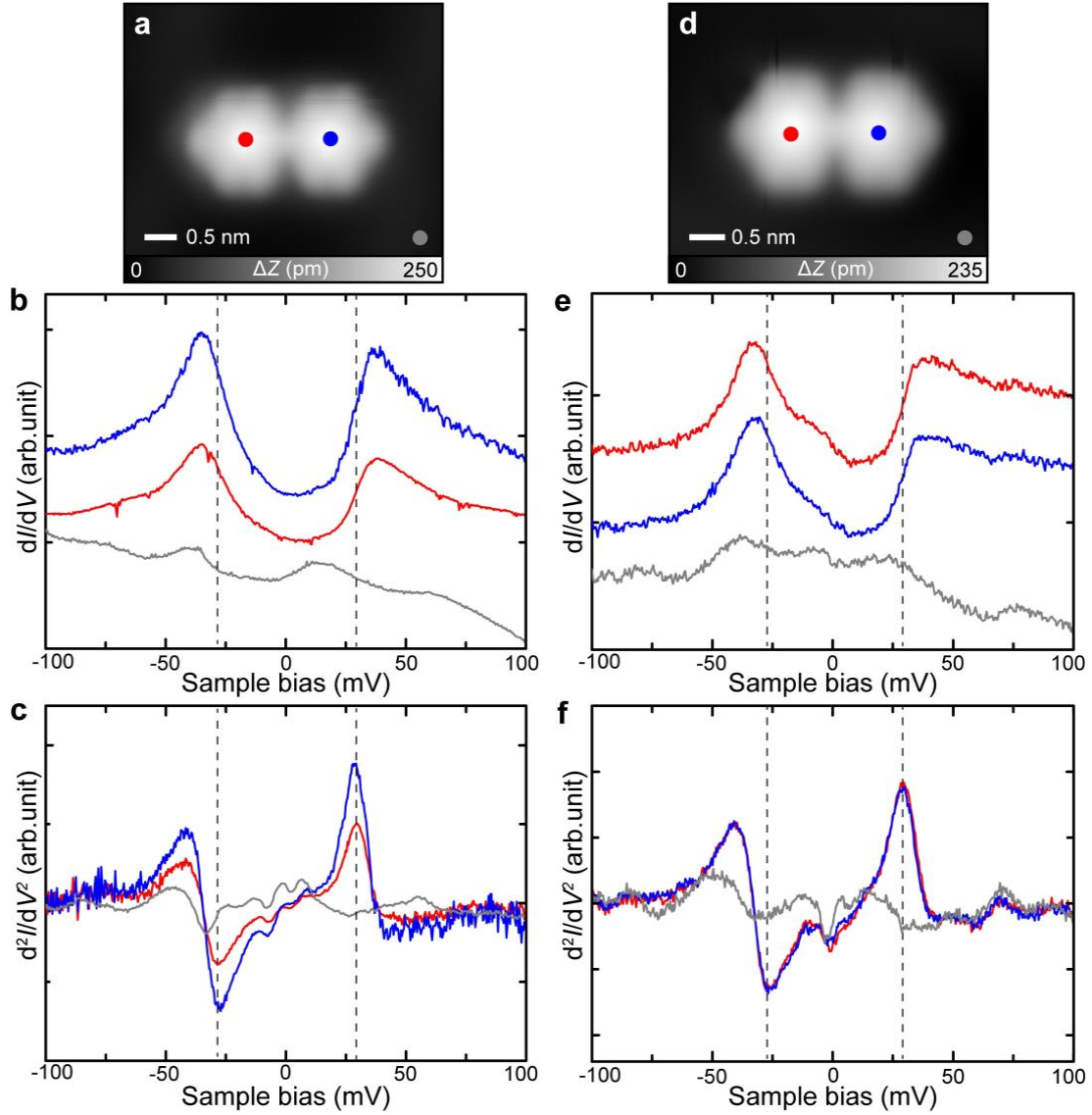

**Fig. S14. Magnetic properties of dimers of N$_2$HBC**. **a**, STM topography. **b,c**, d$I$/d$V$ spectra (**b**) and d$^2I$/d$V^2$ spectra (**c**) recorded on the dimer and a bare Au(111) surface as indicated by colored dots in (**a**). **d**, STM topography of another dimer and **e,f**, the corresponding d$I$/d$V$ spectra (**e**) and d$^2I$/d$V^2$ spectra (**f**). The measured energy of the inelastic spin excitation was 28.5±1 mV. Measurement parameters: $V$ = 200 mV and $I$ = 5 pA in (**a**)(**d**). $V$ = 100 mV and $I$ = 700-1200 pA, $V_{ac}$ = 5 mV in (**b,c**). $V$ = 100 mV and $I$ = 200 pA, $V_{ac}$ = 1 mV in (**e**). $V$ = 100 mV and $I$ = 500 pA, $V_{ac}$ = 5 mV in (**f**).

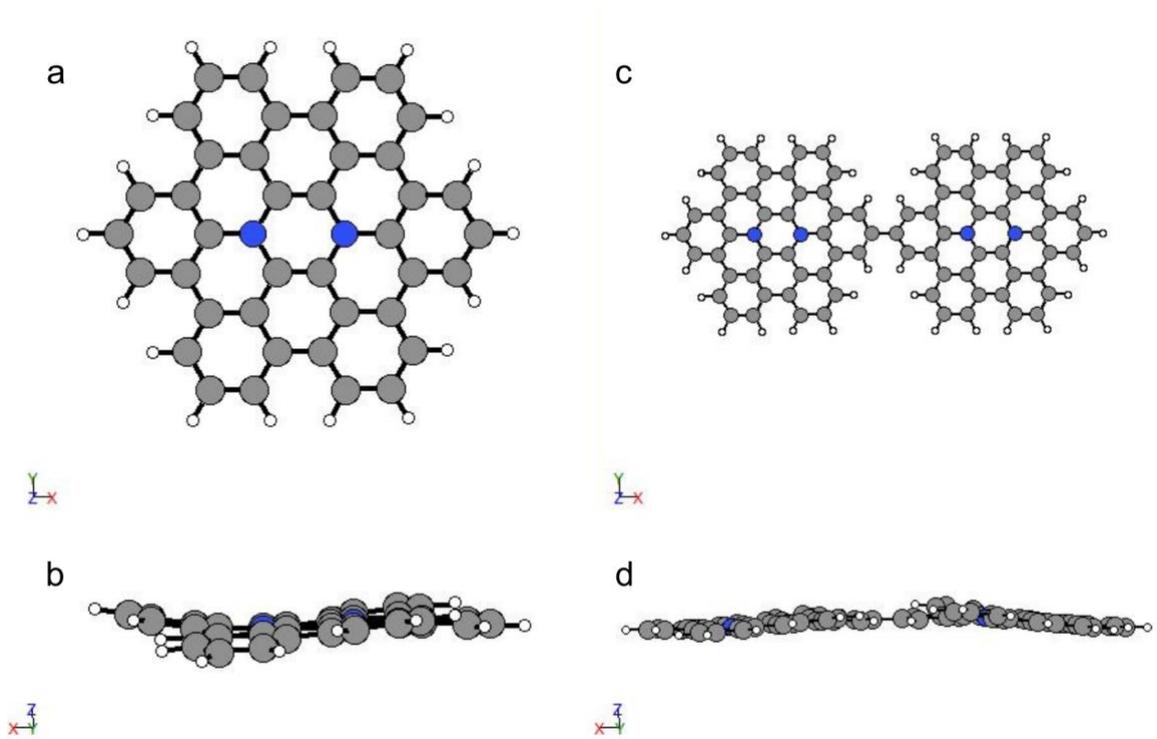

**Fig. S15. a-d**, Top and side views of the free-standing N$_2$HBC$^+$ molecule (**a,b**) and the dimer-N$_2$HBC$^{2+}$ (**c,d**) after relaxation without any constraints.

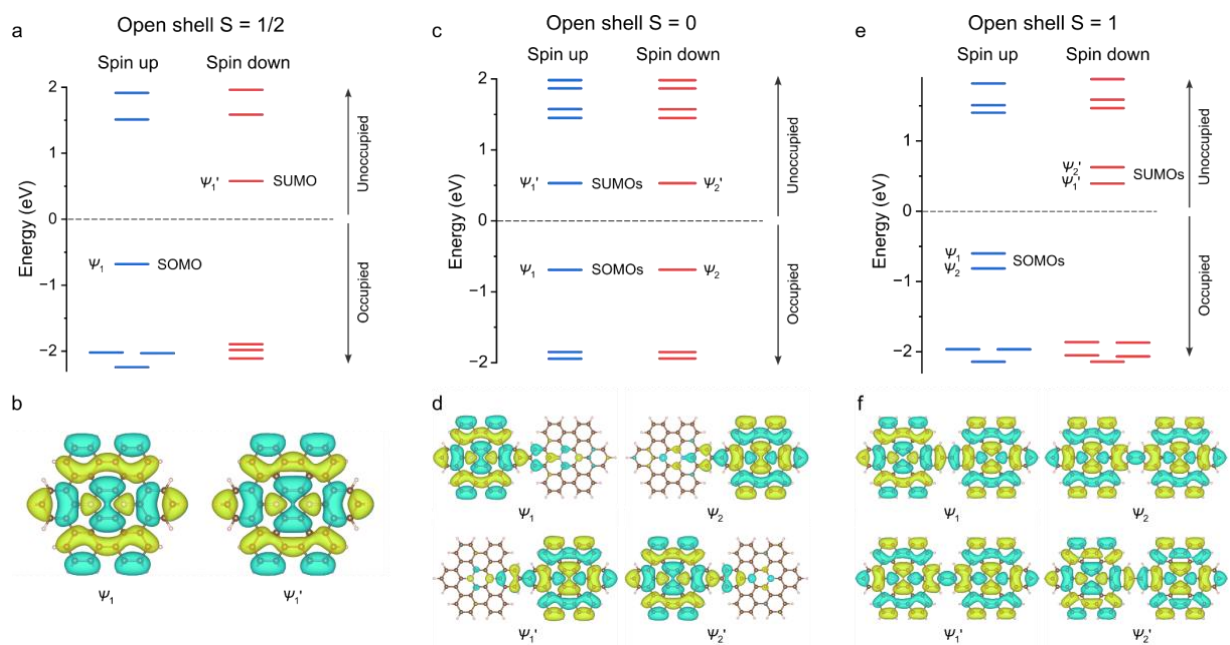

**Fig. S16. Electronic structures of $N_2HBC^+$ molecule and dimer-$N_2HBC^{2+}$. a**, Spin-polarized DFT calculated energy spectrums of the $N_2HBC^+$ molecule. **b**, Spatial distribution of the singly occupied and singly unoccupied molecular orbitals corresponding to the energy levels in (**a**). **c,e,** Spin-polarized DFT calculated energy spectrums of the dimer-$N_2HBC^{2+}$ for its ground singlet state (**c**) and excited triplet state (**e**), respectively. **d,f**, The spatial distributions of the SOMOs (**d**) and LUMOs (**f**) corresponding to the energy levels in (**c**) and (**e**), respectively. The colors in spin distributions: green, spin up; yellow, spin down, and the isosurfaces were set to 0.02 Å$^{-3}$.

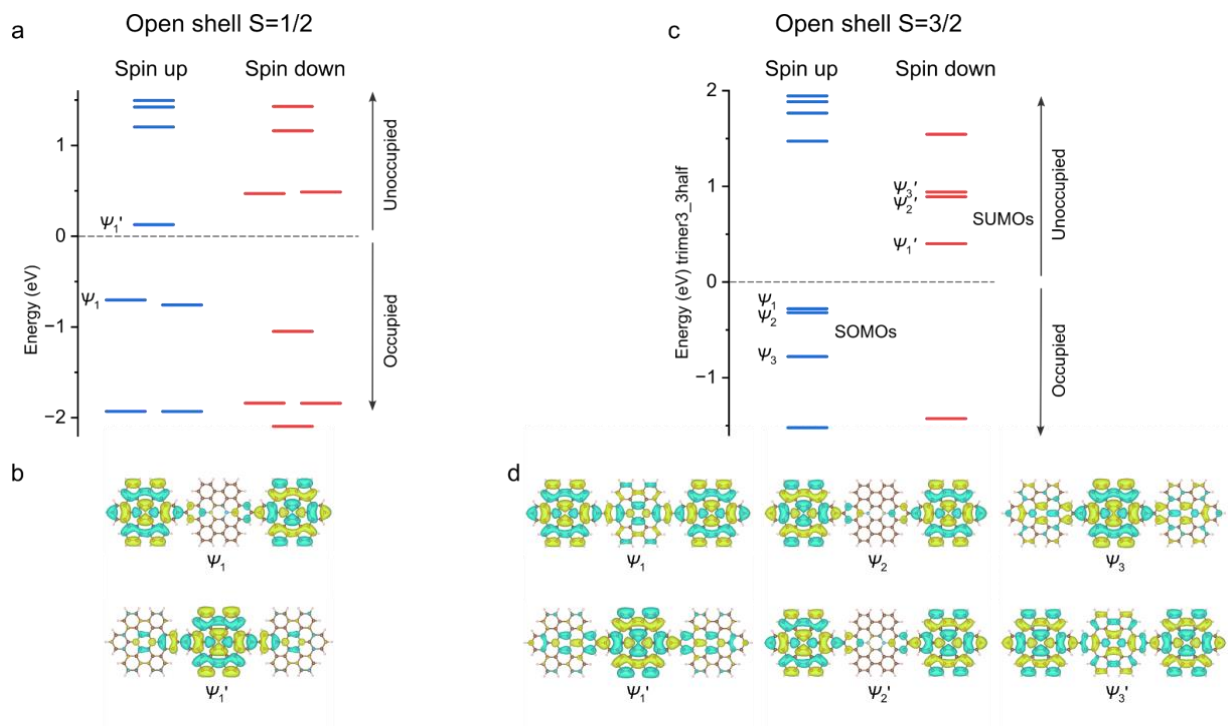

**Fig. S17. Electronic structures of trimer-N$_2$HBC$^{3+}$**. **a,c,** Spin-polarized DFT calculated energy spectrums of the trimer-N$_2$HBC$^{3+}$ for its ground doublet state (**a**) and excited S = 3/2 state (**c**), respectively. **b,d,** Spatial distributions of the SOMOs (**b**) and LUMOs (**d**), among others, corresponding to the energy levels in (**a**) and (**c**), respectively. The colors in spin distributions: green, spin up; yellow, spin down, and the isosurfaces were set to 0.02 Å$^{-3}$.

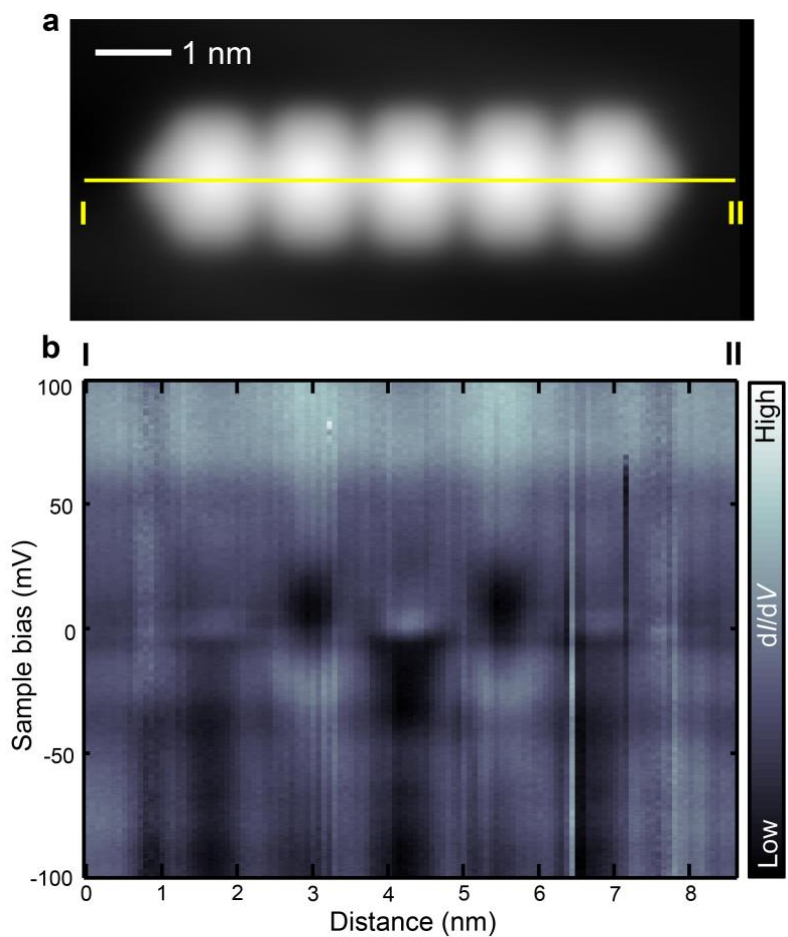

**Fig. S18**. **Magnetic properties of an individual 3 pentamer**. **a**, STM topography. **b**, Two-dimensional d$I$/d$V$ maps of the pentamer, taken along I-II line. Measurement parameters: $V$ = 200 mV and $I$ = 10 pA in (**a**). $V$ = 100 mV, $I$ = 300 pA, $V_{ac}$ = 1 mV and $f$ = 510 Hz in (**b**).

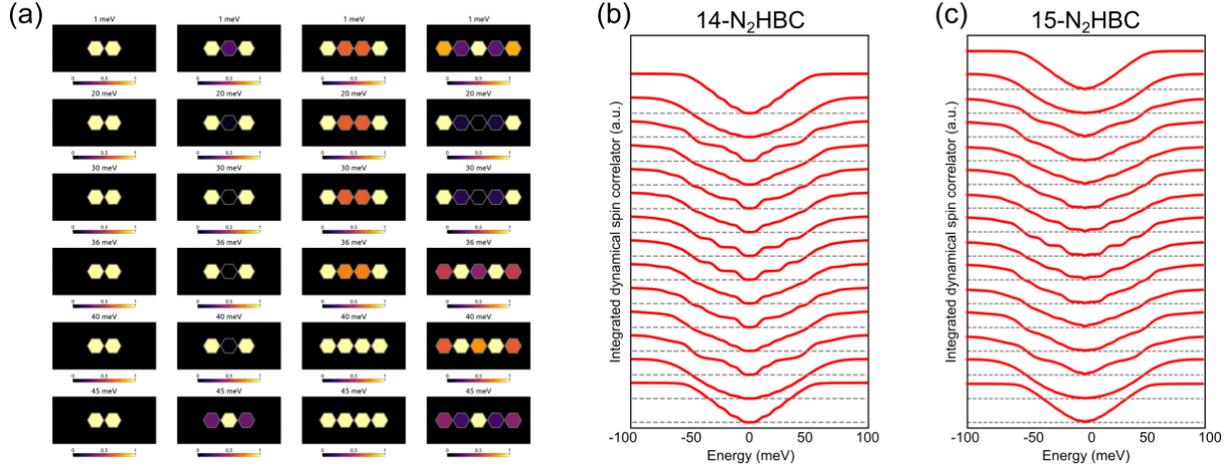

**Fig. S19. DSC maps at different energies and the Heisenberg model applied to longer chains a**, Simulated d$I$/d$V$ normalized maps from dimer to pentamer obtained using our low energy model at several different energies. **b,c**, Integrated dynamical spin correlator corresponding to 14- (**b**) and 15-N$_2$HBC chains (**c**), showing that the Kondo peaks are quenched regardless of the parity of longer chains.

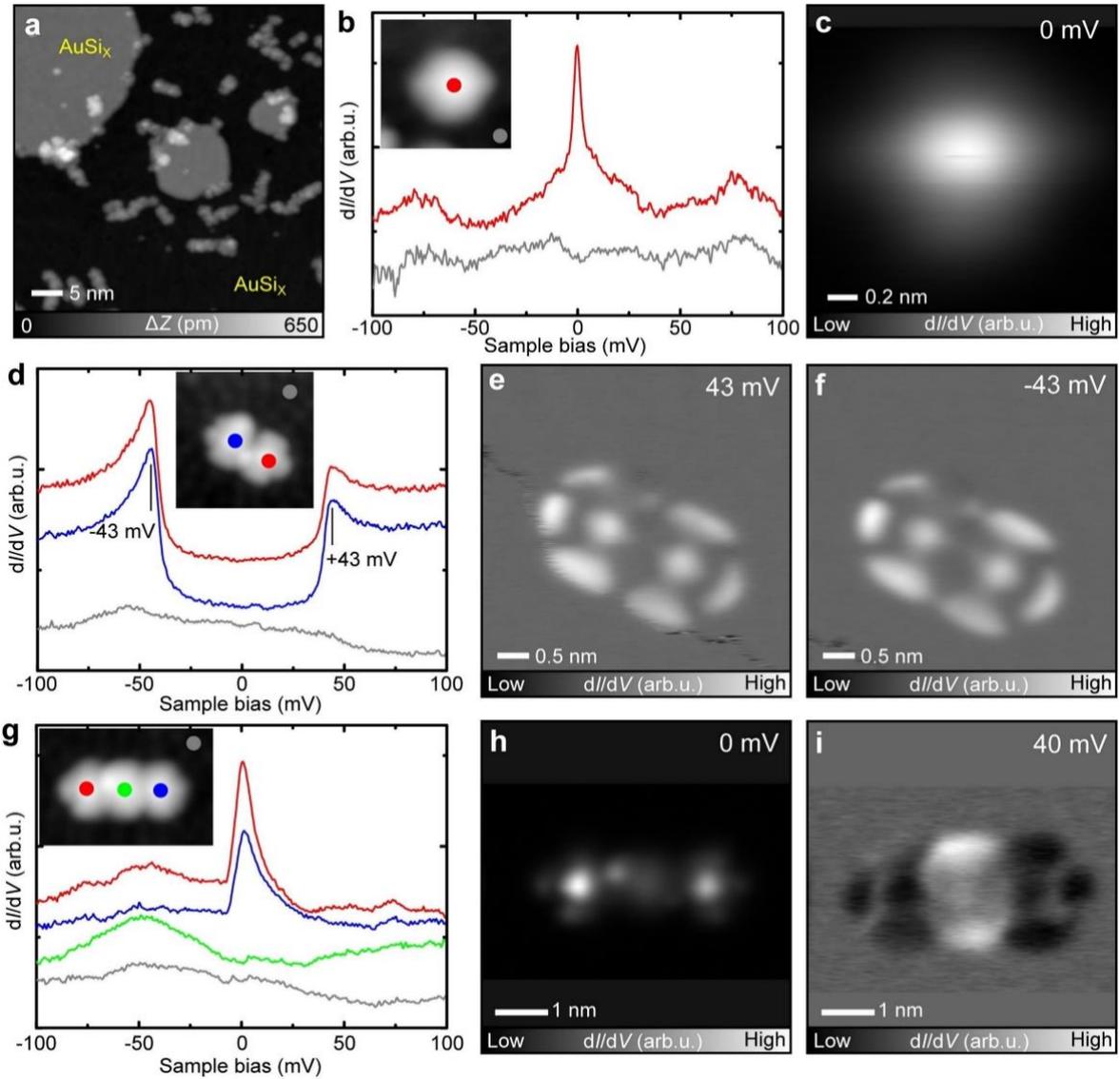

**Fig. S20. Magnetic properties of monomer, dimer and trimer 3 on a AuSi$_X$ layer formed on Au(111). a**, STM topography after depositing Si atoms on the substrate kept at 150 °C. **b**, d$I$/d$V$ curves measured at different sites as indicated by colored dots. Inset shows one monomer on the AuSi$_X$/Au(111) surface. **c**, Constant height d$I$/d$V$ map measured above the monomer at sample bias voltage of 0 V. **d**, d$I$/d$V$ curves measured at different sites as indicated by colored dots. Inset shows one dimer on the AuSi$_X$/Au(111) surface. **e,f,** Constant current d$I$/d$V$ maps measured above the dimer at sample bias voltages of 43 mV, -43 mV, respectively. **g**, d$I$/d$V$ curves measured at different sites as indicated by colored dots. Inset shows one trimer on the AuSi$_X$/Au(111) surface. **h**, Constant height d$I$/d$V$ map measured above the trimer in (**g**) at sample bias voltage of 0 V. **i**, Constant current d$I$/d$V$ map measured above the trimer at sample bias voltages of 40 mV. Measurement parameters: $V$ = 300 mV and $I$ = 1.8 pA in (a).

**Table S1.** Vacuum energy ($E_{vac}$), Fermi level ($E_F$) and work functions ($\phi_{wf} = E_{vac} - E_F$) calculated using PBE for the $N_2HBC$ molecule, dimer-$N_2HBC$ and trimer-$N_2HBC$. The reference value used for Au(111) was also calculated with PBE.

|  | $N_2HBC$ molecule | | dimer-$N_2HBC$ | | trimer-$N_2HBC$ | | Au(111) |
|---|---|---|---|---|---|---|---|
|  | neutral | 1+ charge | neutral | 2+ charge | neutral | 3+ charge | - |
| $E_{vac}$ | -0.073 | 0.104 | -0.135 | 0.170 | -0.146 | 0.205 | - |
| $E_F$ | -3.245 | -5.539 | -3.336 | -5.511 | -3.246 | -5.626 | - |
| $\phi_{wf}$ | **3.17** | **5.64** | **3.20** | **5.68** | **3.10** | **5.83** | **5.19**[S8] |